\documentclass[journal,comsoc]{IEEEtran}

\usepackage{graphicx}
\graphicspath{{Figures/}}

\long\def\comment#1{}

\newfont{\bbb}{msbm10 scaled 700}

\newfont{\bb}{msbm10 scaled 1100}
\newcommand{\CC}{\mbox{\bb C}}

\newcommand{\RR}{\mbox{\bb R}}

\newcommand{\EE}{\mbox{\bb E}}

\newcommand{\Cc}{{\cal C}}

\usepackage{float}
\usepackage{array}
\usepackage{mathtools}
\usepackage{amsmath} 
\usepackage{amssymb}
\usepackage{amsxtra}
\usepackage{amstext}
\usepackage{dsfont} 
\usepackage{bbm}
\usepackage{latexsym}
\usepackage{epsfig}
\usepackage{epstopdf}

\usepackage{caption}
\usepackage{bbm}
\usepackage{enumitem}
\usepackage{multirow}
\usepackage{diagbox}
\usepackage{slashbox}
\usepackage{algorithm}
\usepackage{algpseudocode}
\usepackage{longtable}

\makeatletter
\def\BState{\State\hskip-\ALG@thistlm}
\makeatother

\usepackage{cite}
\usepackage[T1]{fontenc}
\usepackage{setspace}
\usepackage{lscape}
\usepackage[cmintegrals]{newtxmath}
\usepackage{url}
\usepackage{tabularx,colortbl}
\usepackage[table]{xcolor}
\usepackage{units}
\usepackage{cases}
\usepackage{hyperref}
\usepackage[ subrefformat=parens,labelformat=parens, caption=false, font=footnotesize]{subfig}

\usepackage{color}

\newcommand{\nth}[1]{{#1}{\text{th}}}
\newcommand{\abs}[1]{\left|{#1}\right|}

\newcommand{\diagg}[1]{\mathrm{diag}(#1)}
\newcommand{\Hermb}[1]{\mathbf{#1}^\mathrm{H}}

\newcommand{\mbf}[1]{\mathbf{#1}}
\newcommand{\Hr}{\mathrm{H}}
\newcommand{\Tr}{\mathrm{T}}

\newcommand{\thh}{\mathrm{th}}
\newcommand{\CP}{\mathrm{CP}}
\newcommand{\coh}{\mathrm{coh}}
\newcommand{\rms}{\mathrm{rms}}

\newcommand{\maxx}{\mathrm{max}}

\newcommand{\RF}{\mathrm{RF}}
\newcommand{\BBand}{\mathrm{BB}}

\newcommand{\txx}{(\mathrm{t})}
\newcommand{\rxx}{(\mathrm{r})}
\newcommand{\eqlu}{(\mathrm{eq})}
\newcommand{\tx}{\mathrm{tx}}
\newcommand{\rx}{\mathrm{rx}}
\newcommand{\samp}{\mathrm{s}}

\newcommand{\srm}{\mathrm{st}}
\newcommand{\trm}{\mathrm{t}}
\newcommand{\usf}{\mathrm{u}}

\newcommand{\tfd}{\mathrm{TF}}
\newcommand{\ddd}{\mathrm{DD}}

\newcommand{\tot}{\mathrm{tot}}
\newcommand{\vect}[1]{\mathrm{vec}(#1)}

\newcommand{\prob}[1]{\mathrm{Pr}\left(#1\right)}

\newcommand{\FFT}{\mathrm{FFT}}

\makeatletter
\def\ps@IEEEtitlepagestyle{%
  \def\@oddfoot{\mycopyrightnotice}%
  \def\@oddhead{\hbox{}\@IEEEheaderstyle\leftmark\hfil\thepage}\relax
  \def\@evenhead{\@IEEEheaderstyle\thepage\hfil\leftmark\hbox{}}\relax
  \def\@evenfoot{}%
}

\def\mycopyrightnotice{%
  \begin{minipage}{\textwidth}
  \centering \scriptsize
  Copyright~\copyright~2022 IEEE. Personal use of this material is permitted. Permission from IEEE must be obtained for all other uses, in any current or future media, including\\reprinting/republishing this material for advertising or promotional purposes, creating new collective works, for resale or redistribution to servers or lists, or reuse of any copyrighted component of this work in other works by sending a request to pubs-permissions@ieee.org.
  \end{minipage}
}
\makeatother

\begin{document}
\title{Single- versus Multicarrier Terahertz-Band Communications: A Comparative Study}
\author{Simon~Tarboush,
        Hadi~Sarieddeen,~\IEEEmembership{Member,~IEEE,}
        Mohamed-Slim~Alouini,~\IEEEmembership{Fellow,~IEEE}
        and~Tareq~Y.~Al-Naffouri,~\IEEEmembership{Senior Member,~IEEE}
\thanks{
S.\ Tarboush is with Telecommunication Department, Higher Institute for Applied Sciences and Technology (HIAST), Damascus, Syria (e-mail: simon.w.tarboush@gmail.com).
The rest of the authors are with the Department of Computer, Electrical and Mathematical Sciences and Engineering (CEMSE), King Abdullah University of Science and Technology (KAUST), Thuwal, Makkah Province, Kingdom of Saudi Arabia, 23955-6900 (e-mail: hadi.sarieddeen@kaust.edu.sa; slim.alouini@kaust.edu.sa; tareq.alnaffouri@kaust.edu.sa).

This publication is based upon work supported by the King Abdullah University of Science and Technology (KAUST) Office of Sponsored Research (OSR) under Award No. ORA-CRG2021-4695.
}
}
\maketitle

\vspace{-0.7in}

\begin{abstract}
The prospects of utilizing single-carrier (SC) and multi-carrier (MC) waveforms in future terahertz (THz)-band communication systems remain unresolved. On the one hand, the limited multi-path components at high frequencies result in frequency-flat channels that favor low-complexity wideband SC systems. On the other hand, frequency-dependent molecular absorption and transceiver characteristics and the existence of multi-path components in indoor sub-THz systems can still result in frequency-selective channels, favoring off-the-shelf MC schemes such as orthogonal frequency-division multiplexing (OFDM). Variations of SC/MC designs result in different THz spectrum utilization, but spectral efficiency is not the primary concern with substantial available bandwidths; baseband complexity, power efficiency, and hardware impairment constraints are predominant. This paper presents a comprehensive study of SC/MC waveforms for THz communications, utilizing an accurate wideband THz channel model and highlighting the various performance and complexity trade-offs of the candidate schemes.
Simulations demonstrate that discrete-Fourier-transform spread orthogonal time-frequency space (DFT-s-OTFS) achieves a lower peak-to-average power ratio (PAPR) than OFDM and OTFS and enhances immunity to THz impairments and Doppler spreads, but at an increased complexity cost. Moreover, DFT-s-OFDM is a promising candidate that increases robustness to THz impairments and phase noise (PHN) at a low PAPR and overall complexity.

\end{abstract}

\begin{IEEEkeywords}
THz Communications, CP-OFDM, SC-FDE, DFT-s-OFDM, OQAM/FBMC, OTFS, DFT-s-OTFS. 
\end{IEEEkeywords}

\section{Introduction}
\label{sec:intro}

\IEEEPARstart{T}{he} successful deployment of millimeter-wave (mmWave) communications \cite{He9540882} has encouraged researchers to explore the last piece of available spectrum, the terahertz (THz) band over $\unit[0.3\!-\!10]{THz}$, which promises to be an essential ingredient of future ultra-broadband wireless communications \cite{akyildiz2014terahertz,elayan2019terahertz}. Moving towards beyond-fifth generation (B5G) and sixth-generation (6G) wireless networks \cite{dang2020should,rajatheva2020white}, a plethora of services are expected to be supported \cite{zhang20196g}, such as ultra-low latency communications, ubiquitous connectivity, and very high data rates (up to several terabits-per-second (Tbps)). Such features can be leveraged in novel use cases in fixed radio links, wireless local area networks, nano cells, or inter-chip communications. Furthermore, accurate localization, sensing, and imaging applications are promised in the THz band \cite{rappaport2019wireless,sarieddeen2020next}. However, researchers should first overcome several challenges in THz materials and technologies (photonic and electronic) and the corresponding system designs and hardware complexity \cite{sengupta2018terahertz,sarieddeen2020overview}.

The THz-band channel's peculiarities (frequency/distance-dependency and sparsity) impose challenging constraints on the physical layer of future wireless standards. THz signals suffer from severe path loss, which limits the transmission distances to a few meters \cite{jornet2011channel}.
However, long distance sub-THz communications (over hundreds of meters) are still feasible with high-gain antenna arrays \cite{gougeon2020assessment}.
The frequency-selective molecular absorption further results in distance-dependent spectrum fragmentation and shrinking (variable-bandwidth transmission windows) \cite{akyildiz2018combating}. Hence, ultra-massive multiple-input multiple-output (UM-MIMO) antenna arrays and intelligent reflecting surfaces (IRSs) are essential for extending the THz communication range \cite{akyildiz2018combating,faisal2019ultra,sarieddeen2019terahertz}. Furthermore, since the line-of-sight (LoS) path dominates THz-band signal propagation, THz channels tend to be flat-fading. However, a few multi-path components might persist, especially in indoor scenarios, resulting in frequency-selective channels (FSCs) of coherence bandwidths of hundreds of megahertz (MHz) over medium communication distances \cite{han2014multi}. Therefore, THz multi-carrier (MC) schemes retain scenario-specific benefits.

Designing efficient THz-specific waveforms is crucial for unleashing the THz-band's true capabilities. Because bandwidth and spectral efficiency (SE) are not yet a THz bottleneck; low complexity, robustness to hardware impairments and Doppler spreads, and high power efficiency are prioritized. The first sub-THz standard (IEEE 802.15.3d \cite{8066476}) supports switched point-to-point connectivity with data rates exceeding $\unit[100]{Gbps}$, offering two modes: (1) single-carrier (SC) modulation (long-range; high-rate) and (2) on-off keying (OOK) (low-complexity; short-range). OOK utilizes femtosecond-long pulses that could span an ultra-wideband THz spectrum~\cite{Jornet6804405}. However, temporal broadening \cite{han2014multi} and the challenging synchronization procedure question the feasibility of pulse-based modulation.
IEEE 802.15.3d-compliant waveforms are proposed in \cite{shehata2021ieee}, where novel pulse-shaping designs reduce out-of-band (OOB) emissions. Several other projects revisit the physical layer for future B5G sub-THz systems. Most notably, the BRAVE project \cite{dore2018above} advocates for modified SCs schemes, such as continuous phase modulated single-carrier frequency-division multiple-access (CPM SC-FDMA), constrained envelope CPM-SC, differential modulation (like differential phase-shift keying), SC with optimized polar modulation (robust to phase noise (PHN)) \cite{bicais2020optimized}, and variations of spatial- and index-modulation \cite{saad2020generalized,saad2020novel,saad2021novel}.
Block-based SC waveforms, such as discrete-Fourier-transform spread OFDM (DFT-s-OFDM) \cite{sahin2016flexible} can also be investigated.

A variety of THz MC schemes can be explored. In the simplest form, multiple (quasi)-orthogonal non-overlapping SC modulations can be combined with some form of carrier aggregation \cite{yuan2020hybrid}. Cyclic-prefix orthogonal frequency-division multiplexing (CP-OFDM) is well investigated, but it is discouraged at THz \cite{han2016multidamc,hossain2019hierarchical} due to its strong spectral leakage (high OOB emissions), unfavorable peak-to-average power ratio (PAPR) properties (limitations in state-of-the-art THz power amplifiers (PAs) \cite{wang2021power}), strict synchronization procedures, and high sensitivity to Doppler spread. Other MC schemes such as novel fifth-generation new-radio (5G-NR) filter-based candidates have their prospects and challenges. Such filtering is on the whole band in filtered-OFDM (f-OFDM) \cite{abdoli2015filtered}, per-subband (a set of contiguous subcarriers) in universal filtered multi-carrier (UFMC) \cite{vakilian2013universal}, or per-subcarrier in offset quadrature amplitude modulation-based filter-bank multi-carrier (OQAM/FBMC) \cite{bellanger2010fbmc} and generalized frequency-division multiplexing (GFDM) \cite{michailow2014generalized}. Although filter-based schemes overcome some CP-OFDM limitations, reducing OOB emissions and enhancing SE, their high PAPR characteristics and increased implementation complexity can be prohibitive in Tbps baseband systems. For example, the single-tap equalizer is no longer sufficient with CP-free OQAM/FBMC, requiring more complex equalization. Other works propose windowed overlap-and-add OFDM (WOLA-OFDM) \cite{zayani2016wola}, or combinations such as OQAM/GFDM \cite{tarbouche2020performance}.

THz-specific multiple-access techniques are also emerging, such distance-adaptive MCs \cite{han2016multidamc}, hierarchical-bandwidth modulations \cite{hossain2019hierarchical}, and distance-/frequency-dependent adaptive CP-OFDM \cite{boulogeorgos2018distance}, which optimize distance-dependent spectral window utilization.
Other works, such as~\cite{gao2020distance}, develop a novel distance-adaptive absorption peak modulation mainly for THz covert communications by exploiting the unique properties of the THz spectrum (frequency-dependent molecular absorption) through dynamically modulating signals under the molecular absorption peaks. Moreover, the work in ~\cite{shafie2021spectrum} focuses on the multi-band-based spectrum allocation with adaptive sub-band bandwidth to improve the SE of MC-enabled multi-user THz communications, where sub-bands with unequal bandwidths can be assigned to the users. 
Spatial-spread orthogonal frequency-division multiple-access (SS-OFDMA) is another THz MC candidate that realizes frequency-based beam spreading by allocating subcarriers for users in different directions \cite{zhai2021ss}. Similarly, beam-division multiple-access (BDMA) \cite{you2017bdma} schedules mutually non-overlapping beam subsets for users, followed by relaxed per-beam synchronization. Moreover, THz-band non-orthogonal multiple access (NOMA) techniques are argued to be feasible, despite the narrow beams that make user clustering difficult \cite{sarieddeen2021terahertz}. Other conventional techniques that improve SE at a reduced power cost, PAPR, and transceiver complexity, are also being studied for THz communications, including spatial \cite{sarieddeen2019terahertz} and index modulation \cite{loukil2019terahertz} paradigms.

Other novel waveforms target specific THz use cases and constraints. For instance, zero-crossing modulation \cite{fettweis2019zero} uses temporal oversampling and 1-bit quantization to relax hardware requirements, such as in the digital-to-analog converter (DAC) and analog-to-digital converter (ADC). Furthermore, orthogonal time-frequency space (OTFS) waveform \cite{hadani2017orthogonal} is tailored for time-variant (TV) channels and high Doppler spreads, which arise in high-speed THz communication scenarios such as vehicle-to-everything (V2X), drone, and ultra-high-speed rail communications. OTFS is superior in block error rate performance to CP-OFDM when assuming mmWave LoS V2X channels \cite{wiffen2018comparison}; also when accounting for oscillator PHN impairments \cite{surabhi2019otfs}.
To meet the THz integrated sensing and communication (ISAC) requirements, the utilization of DFT-s-OFDM, with some modifications, is discussed in~\cite{wu2021sensing}. Most recently, a novel scheme called DFT-s-OTFS is proposed \cite{wu2021dft,wu2022dft} to address the severe Doppler effects and PAPR challenges of THz ISAC.

Many performance metrics need to be considered when designing THz waveforms, such as bit error rate (BER), PAPR, and baseband computational complexity. Furthermore, hardware imperfections and radio frequency (RF) impairments critically impact THz waveform design, where candidate THz materials/hardware are still under development. Hardware imperfections include PA non-linearity, wideband in/quadrature-phase imbalance (IQI) \cite{sha2021channel}, phase uncertainty in the phase-shifters (PSs) \cite{lin2015indoor}, and PHN (studied for SC schemes \cite{bicais2019phase} and CP-OFDM \cite{neshaastegaran2020effect} in sub-THz and THz \cite{sha2021channel} systems). THz channel-induced phenomena such as beam split and misalignment \cite{tarboush2021teramimo} are also critical, especially with UM-MIMO systems.
Moreover, synchronization becomes more challenging with carrier frequency offset (CFO) and symbol-timing offset (STO) at THz frequencies.
Subcarrier spacing (SCS), its impact on PHN, and the design of phase-tracking reference signals are studied in \cite{levanen2020mobile,tervo20205g} to assess whether CP-OFDM and DFT-s-OFDM can support mmWave and sub-THz communications. Moreover, a THz SC frequency-domain equalization technique (SC-FDE) is developed in \cite{sha2021channel}, and a pilot design strategy based on index modulation is proposed in \cite{mao2021terahertz}. SC systems are found superior to CP-OFDM in mmWave systems \cite{buzzi2017single} when taking into account the transmitter PA  non-linearities. For indoor THz scenarios, SC-FDMA with linear equalization is shown to be superior to CP-OFDM and SC with linear-/decision-feedback-equalization \cite{schram2020comparison}.

The literature lacks a holistic and fair comparative study of THz-band SC/MC schemes, and this work attempts to fill this gap. The main aim is to analyze a plethora of candidate waveforms to draw recommendations on the suitable waveforms for specific THz use cases. The main contributions of this paper are summarized as follows:
\begin{itemize}
  \item Studying the THz compatibility of multiple waveforms, namely, SC-FDE, CP-OFDM, DFT-s-OFDM, OTFS, DFT-s-OTFS, and OQAM/FBMC, adopting our newly developed accurate THz channel model/simulator (TeraMIMO \cite{tarboush2021teramimo}).
  \item Analyzing normalized SE, transmit time interval (TTI) (a delay component of the physical layer latency), OOB emissions (reference THz IEEE 802.15.3d spectral mask), PAPR (theoretical bounds), and computational complexity.
  \item Providing a fair comparison of waveforms under THz-specific scenarios such as oscillator PHN (by studying a Gaussian uncorrelated PHN model), mobility, and beam split.
  \item Promoting DFT-s-OFDM and DFT-s-OTFS as promising schemes for future B5G/6G networks.
\end{itemize}
The remainder of this paper is organized as follows: Sec.~\ref{sec:system_ch_model} first introduces the system and channel models. Then, Sec.~\ref{sec:sc_mc_kpi} presents several key performance indicators (KPIs) to compare different waveforms. Sec.~\ref{sec:sc_mc_descr} details a general framework for analyzing the studied SC/MC waveforms. Afterward, extensive simulation results validate our analyses in Sec.~\ref{sec:sim_res}, where recommendations of suitable waveforms for specific scenarios are introduced. Sec.~\ref{sec:conclusion_future_pros} concludes the paper.
Regarding notation, non-bold lower case, bold lower case, and bold upper case letters correspond to scalars, vectors, and matrices: $a[n]$ denotes the $n\thh$ element of $\mbf{a}$ and $\mbf{a}[m]$ and $a[n,m]$ denote the $m\thh$ column and the $(n,m)\thh$ element of $\mbf{A}$, respectively.
$\mbf{I}_N$ is the identity matrix of size N, $\mbf{0}_{N,M}$ is a zero matrix of size $\mathrm{N}\!\times\!\mathrm{M}$, and $\mbf{a}_N$ is a vector of size $\mathrm{N}$.
The superscripts ${(\cdot)}^\Tr$, ${(\cdot)}^\ast$, ${(\cdot)}^\Hr$, ${(\cdot)}^\mathrm{-1}$, and ${(\cdot)}^n$ stand for the transpose, conjugate, conjugate transpose, inverse, and $n\thh$-power functions, respectively. $\abs\cdot$ is the absolute value (or set cardinality), $\diagg{\!a_0,a_1,\dots,a_{N-1}\!}$ is an $\mathrm{N}\!\times\!\mathrm{N}$ diagonal matrix of diagonal entries $\!a_0,a_1,\dots,a_{N-1}\!$, $\vect{\mbf{A}}$ is the vectorized matrix representation that stacks the columns of $\mbf{A}$ in a single column, $\EE(\cdot)$ is the expectation operator, and $\prob{\cdot}$ is the probability density function. The notations $\otimes$, $[\cdot]_N$, $\langle\,,\rangle$, $\mathcal{R}({\cdot})$, and $j\!=\!\sqrt{-1}$ denote the Kronecker product, remainder modulo $\mathrm{N}$, inner product, real part, and imaginary unit, respectively. The superscripts ${\txx}$ and ${\rxx}$ denote transmitter (Tx) and receiver (Rx) parameters, respectively. $\mathcal{N}(\varphi,\sigma^2)$ is the distribution of a Gaussian random variable of mean $\varphi$ and variance $\sigma^2$, $\mathcal{CN}(\mbf{a},\mbf{\Sigma})$ is the distribution of a complex Gaussian random vector of mean $\mbf{a}$ and covariance matrix $\mbf{\Sigma}$.
The normalized $N$-point DFT and IDFT matrices are denoted by $\mbf{F}_N$ and $\mbf{F}^\Hr_N$, respectively.
The used acronyms are summarized in Tables~\ref{table:AbbreviationsTable1} and \ref{table:AbbreviationsTable2}.

\section{System and Channel Model}
\label{sec:system_ch_model}

\begin{figure}[!t]
   \begin{minipage}{0.48\textwidth}
     \centering
     \includegraphics[width=.99\linewidth]{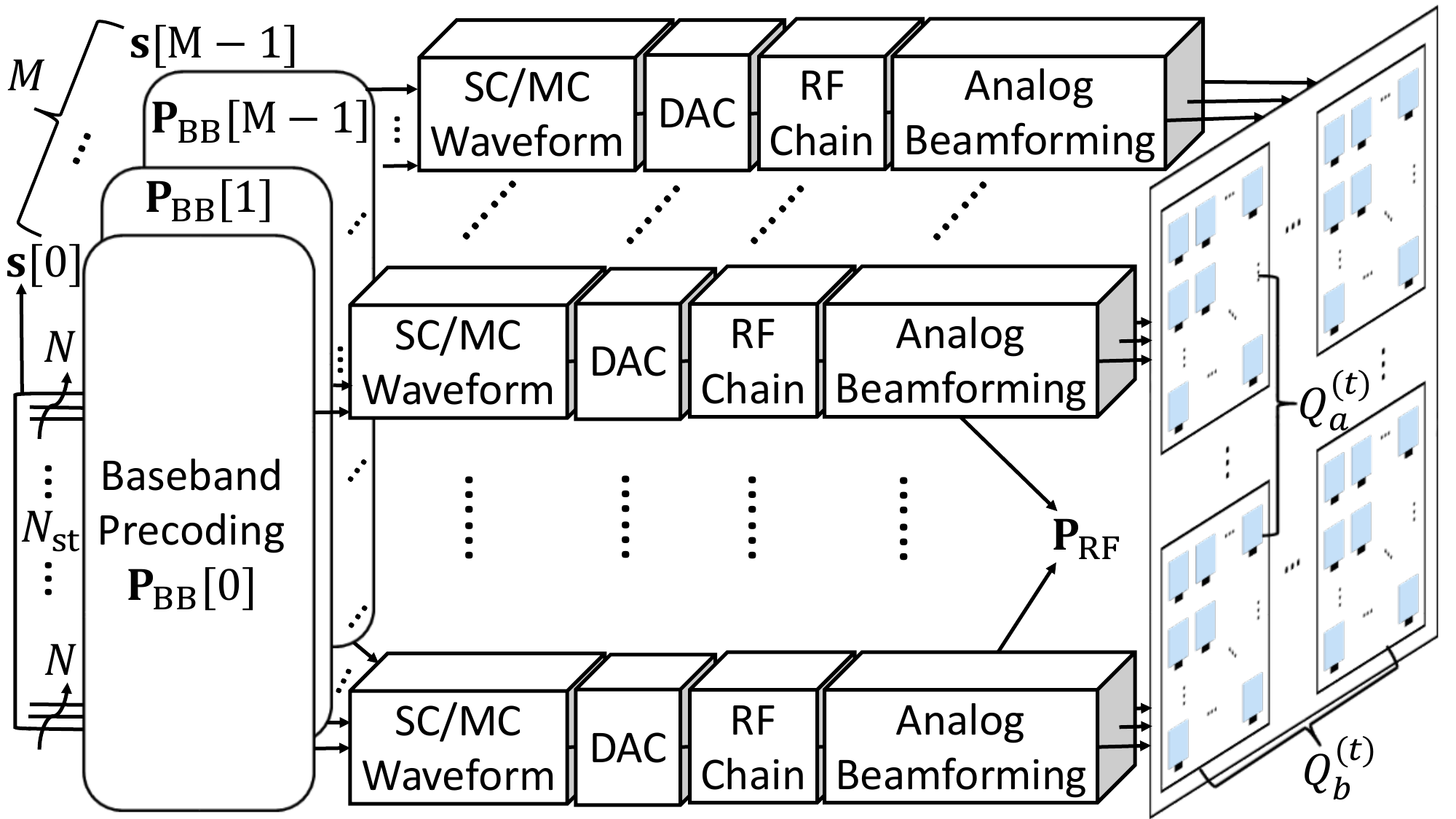}
     \caption{Block diagram of a THz-band UM-MIMO transmitter.}\label{fig:MC_txrx_SysModel}
   \end{minipage}\hfill
   \begin{minipage}{0.48\textwidth}
     \centering
     \includegraphics[width=.99\linewidth]{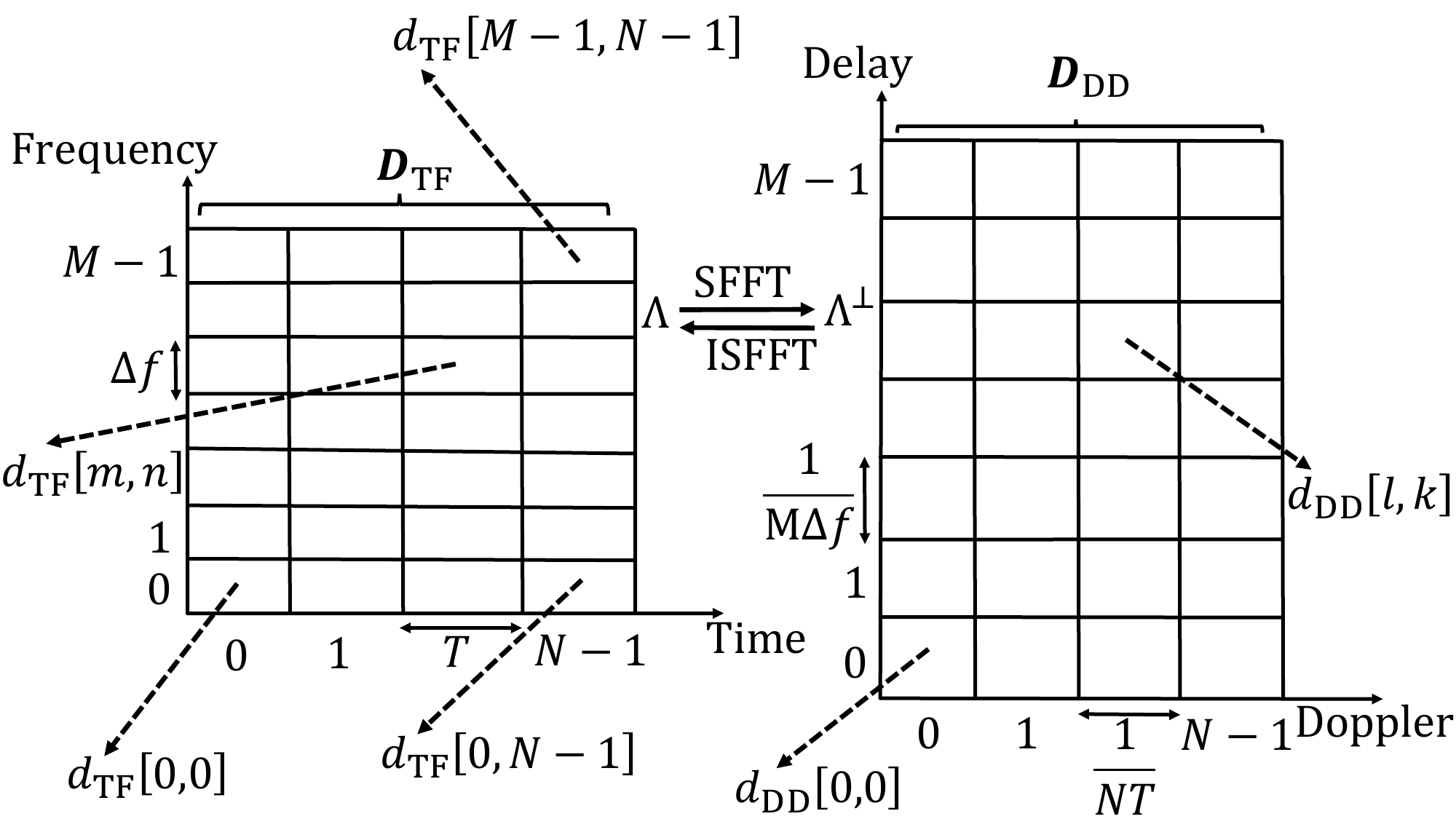}
     \caption{Illustration of time-frequency delay-Doppler lattices.}\label{fig:lattice_TF_DD}
   \end{minipage}
\end{figure}

The main aim of this work is to evaluate the performance of candidate SC/MC waveforms in realistic THz settings, including massive antenna dimensions and ultra-wide bandwidths. We adopt the array-of-subarrays (AoSA) architecture of TeraMIMO \cite{tarboush2021teramimo}, in which each subarray (SA) is composed of many antenna elements (AEs), as depicted in Fig.~\ref{fig:MC_txrx_SysModel}. AoSAs can mitigate THz hardware constraints and combat the limited communication distance problem using low-complexity beamforming \cite{sarieddeen2020overview}.
{The model assumes} $Q\!=\!Q_a\!\times\! Q_b$ SAs, and $\bar{Q}\!=\!\bar{Q}_a\times\!\bar{Q}_b$ tightly-packed directional AEs per SA. Each AE is attached to a wideband THz analog PS of acceptable phase error, return loss, and insertion loss~\cite{lin2015indoor} (such PSs can be implemented using graphene transmission lines in plasmonic solutions \cite{chen2012terahertz}). The AoSAs are assumed to realize sub-connected hybrid beamforming, with analog beamforming over the AEs of each SA. Each RF chain thus drives one disjoint SA, reducing power consumption and complexity; the SAs provide the spatial diversity gain. 

For SCs, {this work considers} SC-FDE, DFT-s-OFDM, and DFT-s-OTFS. For MCs, {the work investigates} CP-OFDM, OQAM/FBMC, and OTFS, assuming $M$-subcarriers. The $\nth{m}$-subcarrier Rx signal is
\begin{equation}
    \Tilde{\mbf{y}}[m]=\mbf{W}^\Hr_{\BBand}[m]\mbf{W}^\Tr_{\RF}\mbf{H}[m]\Tilde{\mbf{x}}[m]+\mbf{W}^\Hr_{\BBand}[m]\mbf{W}^\Tr_{\RF}\mbf{n}[m],
    \label{eq:rx_subc_sig}
\end{equation}
where assuming perfect time and frequency synchronization (no STO or CFO), the received signal is processed using an RF combining matrix, $\mbf{W}_{\RF}\!\in\!\CC^{Q^{\rxx}\bar{Q}^{\rxx}\!\times\!Q^{\rxx}}$, and a digital baseband combining matrix, $\mbf{W}_{\BBand}[m]\!\in\!\CC^{Q^{\rxx}\!\times\!N_\tot}$; $\mbf{n}[m]\!\in\!\CC^{Q^{\rxx}\bar{Q}^{\rxx}\!\times\!1}$ is the additive white Gaussian noise (AWGN) vector of independently distributed $\mathcal{CN}(\mbf{0}_{Q^{\rxx}\bar{Q}^{\rxx}},\sigma_n^2\mbf{I}_{Q^{\rxx}\bar{Q}^{\rxx}})$ elements of noise power $\sigma_n^2$.
Note that $N_\tot\!=\!N_\srm\!\times\! N$, where $N_\srm\!\leq\! Q^{\txx}$ is the number of data streams ($Q^{\txx}$ is also the number of Tx RF chains), and $N$ is the number of MC symbols per frame.

The UM-MIMO channel matrix, $\mbf{H}[m]\!\in\!\CC^{Q^{\rxx}\bar{Q}^{\rxx}\!\times\! Q^{\txx}\bar{Q}^{\txx}}$, represents the overall complex channel at the $\nth{m}$-subcarrier; assuming a time-invariant (TIV)-FSC, $\mbf{H}$ can be expressed as 
\begin{equation}
    \mbf{H}[m]= \begin{bmatrix}
    \mbf{H}_{1,1}[m]& \cdots & \mbf{H}_{1,Q^{(\mathrm{t})}}[m]\\
    \vdots & \ddots & \vdots\\
     \mbf{H}_{Q^{(\mathrm{r})},1}[m] & \cdots & \mbf{H}_{Q^{(\mathrm{r})},Q^{(\mathrm{t})}}[m]\\
    \end{bmatrix},
    \label{eq:overall_H_sub_H}
\end{equation}
where $\mbf{H}_{q^{\rxx},q^{\txx}}[m]\!\in\!\CC^{\bar{Q}^{\rxx}\times\bar{Q}^{\txx}}$ denotes the channel response between the $\nth{q^{\txx}}$ Tx SA and the $\nth{q^{\rxx}}$ Rx SA. Further details on the channel model can be found in \cite{tarboush2021teramimo} and equations therein (Eqs. $(15)$ and $(16)$ define $\mbf{H}_{q^{\rxx},q^{\txx}}[m]\!\in\!\CC^{\bar{Q}^{\rxx}\times\bar{Q}^{\txx}}$ in the delay and frequency domains, respectively).
The discrete-time Tx complex baseband signal at the $\nth{m}$-subcarrier is
\begin{equation}
    \Tilde{\mbf{x}}[m]=\mbf{P}_{\RF}\mbf{P}_{\BBand}[m]\mbf{s}[m],
    \label{eq:tx_subc_sig}
\end{equation}
where $\mbf{P}_{\BBand}[m]\!\in\!\CC^{Q^{\txx}\!\times\!N_\tot}$ is the digital baseband precoding matrix per subcarrier, $\mbf{P}_{\RF}\!\in\!\CC^{Q^{\txx}\bar{Q}^{\txx}\!\times\!Q^{\txx}}$ is the analog RF beamforming matrix, and $\mbf{s}[m]\!=\!{\left[{s_1,s_2,\dots,s_{N_\tot}}\right]}^\Tr \!\in\! \mathcal{X}^{N_\tot\times1}$ is the information-bearing symbol vector consisting of data symbols drawn from a quadrature amplitude modulation (QAM) constellation, $\mathcal{X}$. We assume normalized symbols, $\EE\left(\mbf{s}[m]\mbf{s}^\ast[m]\right)\!=\!\frac{P_\trm}{M\!N_\tot}\mbf{I}_{N_\tot}$, where $P_\trm$ is the average total Tx power over $M$-subcarriers. We adopt this model for simulating THz-specific beam-split effects. For other scenarios, the system model reduces to a single-input single-output (SISO) model. We adapt the TeraMIMO THz channel simulator \cite{tarboush2021teramimo} to account for diverse scenarios.

\section{Key Performance Indicators for SC/MC Waveform Performance Evaluation}
\label{sec:sc_mc_kpi}

Choosing a suitable waveform is a challenging task that depends on several conflicting communication system performance requirements and design criteria. For fairness of comparison, we consider the transmission of $M\!\times\!N$ complex symbols of bandwidth $B\!=\!M\!\Delta f$, with SCS $\Delta f$ and frame duration $T_f\!=\!N\!T$, for both SC and MC schemes; the signal period ($T$) differs between waveforms. The time-frequency (TF) domain is discretized into a lattice, $\Lambda$, by sampling time and frequency at integer multiples of $T$ and $\Delta f$, respectively
\begin{equation}
    \label{eq:lattice_tf}
    \Lambda=\left\{\left(nT,m\Delta f\right),n\!=\!0,\dots,N\!-\!1, m\!=\!0,\dots,M\!-\!1\right\}.
\end{equation}
Similarly, the delay-Doppler (DD) plane is discretized into
\begin{equation}
    \label{eq:lattice_dd}
    \Lambda^\perp=\left\{\left(\frac{k}{NT},\frac{l}{M\Delta f}\right),k\!=\!0,\dots,N\!-\!1, l\!=\!0,\dots,M\!-\!1\right\},
\end{equation}
where $\!\frac{1}{NT},\!\frac{1}{M\Delta f}\!$ define the Doppler and delay domain resolutions, respectively.
The maximum supported Doppler and delay spreads are $\!\nu_\maxx\!=\!\frac{\upsilon}{c}f_c\!<\!1/T\!$ and $\!\tau_\maxx\!\!<\!\!1/\Delta f\!$, respectively, where $\upsilon$ is the user velocity, $c$ is the speed of light, and $f_c$ is the carrier frequency.

Both TF and DD lattices are shown in Fig.~\ref{fig:lattice_TF_DD}, where we denote by $\mbf{D}_\tfd$ and $\mbf{D}_\ddd\!\in\!\CC^{M\!\times\!N}$ the data symbol matrices (of elements $d_\tfd[m,n]$ and ${d_\ddd[l,k]}$) in the TF and DD domains, respectively. In vector form, $\mbf{d}_\tfd\!=\!\vect{\mbf{D}_\tfd}$ and $\mbf{d}_\ddd\!=\!\vect{\mbf{D}_\ddd}$. Furthermore,
$\mbf{d}_M^\tfd\!\in\!\CC^{M\times1}$ is a column of $\mbf{D}_\tfd$ (of elements $d_\tfd[m]$). In the case of DFT-s-OFDM, the data symbol matrix is $\bar{\mbf{D}}_\tfd\!\in\!\CC^{\bar{M}\!\times\!N}$, a sub-matrix of $\mbf{D}_\tfd$, where $\bar{M}$ represents the number of Tx symbols modulated over $M$ subcarriers. We also denote by $\bar{\mbf{d}}_{\bar{M}}^\tfd\!\in\!\CC^{\bar{M}\times1}$ a column of $\bar{\mbf{D}}_\tfd$.
Moreover, for DFT-s-OTFS, the data matrix is $\mbf{\bar{D}}_{\bar{N}\!M}\!\in\!\CC^{M\times\!\bar{N}}$, where $\bar{N}\!M$ represents the number of Tx symbols.
Note that $\mbf{S}\!=\![\mbf{s}[0], \mbf{s}[1], \dots,\mbf{s}[M\!-\!1] ]$ of~\eqref{eq:tx_subc_sig}, for a single data stream ($N_\srm\!=\!1$; no digital baseband precoding), reduces to $\mbf{D}_\tfd^\Tr$.

\begin{figure*}
  \centering
  \captionsetup{justification=centering}
  \includegraphics[width=0.7\textwidth]{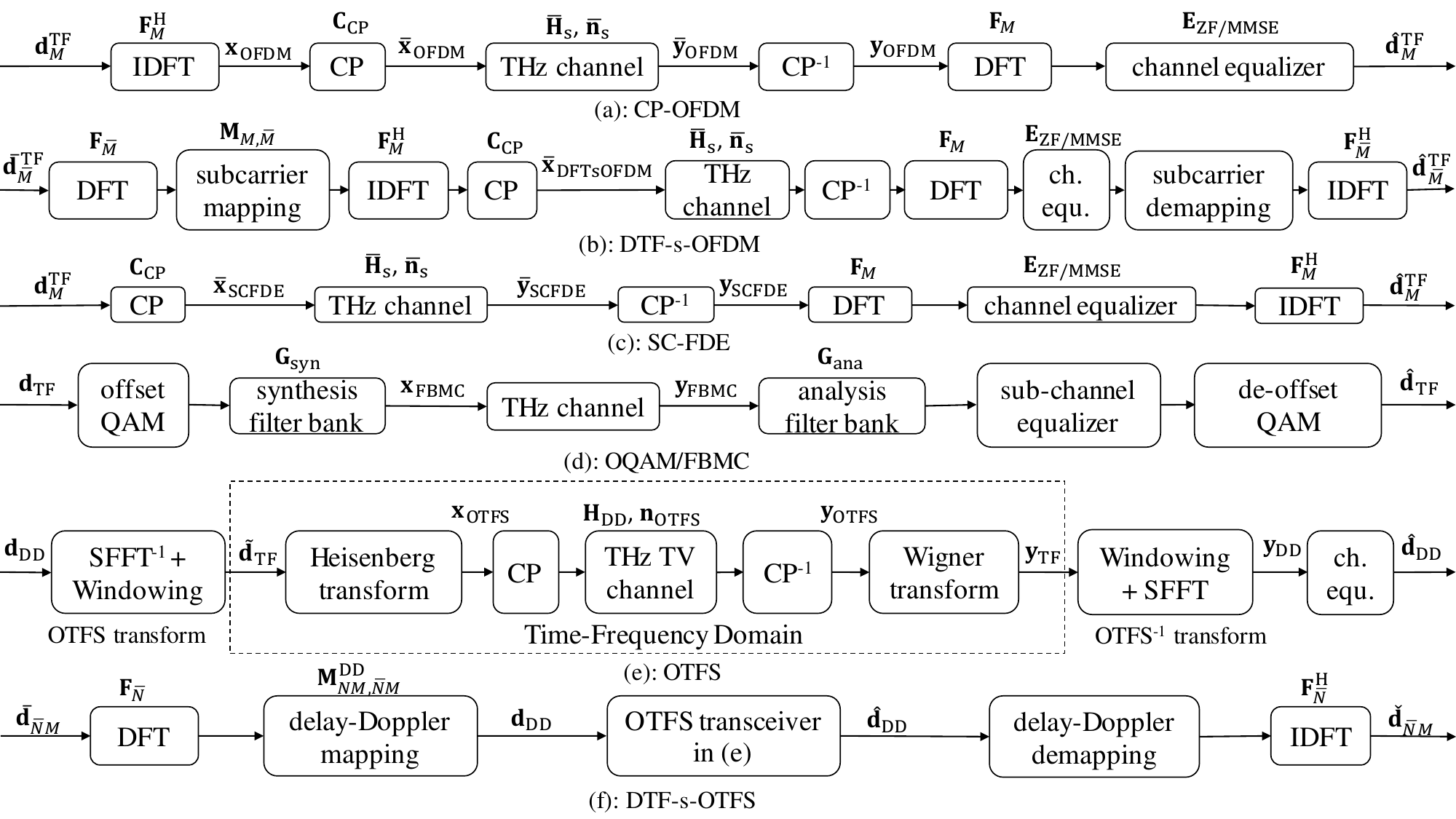}
  \caption{SC/MC transceiver block diagrams: (a) CP-OFDM, (b) DFT-s-OFDM, (c) SC-FDE, (d) OQAM/FBMC, (e) OTFS, and (f) DFT-s-OTFS}.
  \label{fig:SCMC_scheme_block_diagram}
  \vspace{-5mm}
\end{figure*}

The general form of a continuous-time MC modulator, $x(t)$, can be expressed using the discrete Heisenberg transform \cite{hadani2017orthogonal}, parameterized by a pulse-shaping prototype filter, $g_\tx(t)$, as
\begin{align}
    \label{eq:mc_generaltx}
    x(t)&=\sum_{m=0}^{M-1}\sum_{n=0}^{N-1}d_\tfd[m,n]g_{m,n}(t),\\
    g_{m,n}(t)&=g_\tx(t-nT)e^{j2\pi m\Delta f(t-nT)},
    \label{eq:mc_generaltx_gt}
\end{align}
where $d_\tfd[m,n]$ represents the Tx symbol at subcarrier-index $m$ and time-index $n$. The complex orthogonality condition for the basis pulse $g_{m,n}(t)$ is expressed as $\langle\,\!g_{m_1,n_1}(t),g_{m_2,n_2}(t)\!\rangle\!=\!\delta_{(m_2-m_1),(n_2-n_1)}$, with $\delta$ being the Kronecker delta function.
The discrete-time representation of~\eqref{eq:mc_generaltx} (Nyquist sampling at $F_\samp\!=\!\frac{1}{T_\samp}\!=\!B$; limited by ADC/DAC specifications) is
\begin{equation}
    x[uT_\samp]=\sum_{m=0}^{M-1}\sum_{n=0}^{N-1}d_\tfd[m,n]g_{m,n}[uT_\samp],
    \label{eq:mc_generaltx_dt}
\end{equation}
where $u\!=\!\left\{0,1,\dots,M\!N\!-\!1\right\}$. The PAPR of a discrete-time signal $x[u]$ over a finite observation period $N_{\mathrm{per}}$ is expressed as a random variable \cite{chafii2014closed}
\begin{equation}
    \label{eq:PAPR_def}
    \mathrm{PAPR}\left(x[u]\right)=\max_{u\in [0,N_{\mathrm{per}}-1]}\left(\abs{x[u]}^2\right)/\EE\left(\abs{x[u]}^2\right),
\end{equation}
the statistical behavior of which can be estimated through numerical simulations. 
However, the PAPR for the discrete-time baseband signal, $x[u]$, is noticeably lower than the PAPR of the continuous-time baseband signal, $x(t)$. Thus, we perform $L$-times interpolation
(oversampling), where $L\geq4$, to
obtain a close PAPR to that of $x(t)$.
We characterize the complementary cumulative distribution function (CCDF) of PAPR.
In the remainder of this section, we detail various KPIs and introduce several schemes, namely, CP-OFDM, DFT-s-OFDM, SC-FDE, OQAM/FBMC, OTFS, and DFT-s-OTFS, as illustrated in Fig.~\ref{fig:SCMC_scheme_block_diagram}.
\subsection{Spectral Efficiency and Transmit Time Interval Latency}
\label{sec:se_e2e_comp}

The SE (bits/sec/Hz) is an essential indicator of throughput and achievable rate for a given bandwidth.
Since the THz band promises huge available bandwidths, unlike below $\unit[6]{GHz}$ communications, SE is not a primary concern.
However, SE is still important for data demanding use cases, such as THz-enabled holographic video meeting, augmented reality (AR), and virtual reality (VR).
Similarly, the TTI latency, defined as the minimum time to transmit each packet of data \cite{chen2018ultra}, is waveform-dependant (overlapping in OQAM/FBMC lengthens the frame duration, for example).
Nevertheless, the ultra-broadband THz bandwidth ($B$) ensures a very small sampling period ($T_\samp$).
Note that physical layer latency includes other delay components~\cite{chen2018ultra}, such as the signal processing time of the equalizer and channel encoder/decoder (but are not included in our latency definition and computations). Signal processing latency is more critical at THz frequencies and depends on the used waveform.

\subsection{Power Spectral Density and Out-of-Band Emissions}
\label{sec:psd_oob_comp}

The power spectral density (PSD) and OOB emissions follow strict standard regulations to meet spectrum mask requirements. 
For example, the international telecommunications union (ITU) radio regulation $5.340$ prohibits transmissions in ten passive bands over $\unit[100\!-\!252]{GHz}$ to protect deep space observatories and satellite sensors \cite{xing2021terahertz}, resulting in a maximum available contiguous bandwidth of $\unit[23]{GHz}$. OOB emissions are also critical in integrated space-air-ground THz networks. It is thus important to study OOB emissions-induced interference to neighboring systems and among multiple users, highlighting the role of carrier-aggregation techniques. The severity of OOB emissions is dictated by bandwidth, required SE, and neighboring co-operating systems. The waveform Tx spectrums in the IEEE 802.15.3d sub-THz standard are described for different bandwidths in \cite{8066476}.

\subsection{Transceiver Complexity}
\label{sec:txrxcomplexity}

The computational complexity of the studied SC/MC transceivers is arguably the most important KPI to consider, given the limited processing capabilities at Tbps and the need for low-cost and low-power solutions.
Without loss of generality, the computations only consider the number of real multiplications per unit of time in the modulation, demodulation, and equalization processes. The complexity of channel coding and decoding are important in their own right but not included in our study.

The number of real multiplications in an $M$-point fast Fourier transform (FFT)/inverse FFT (IFFT) (split-radix algorithm) is \cite{gerzaguet20175g}
\begin{equation}
    \label{eq:fft_comp}
    \bar{\Cc}_\FFT(M)=M(\log_2{\left(M\right)}-3)+4.
\end{equation}
As illustrated in Fig.~\ref{fig:SCMC_scheme_block_diagram}(a), IFFT/FFT is followed by rectangular pulse-shaping in CP-OFDM, resulting in a complexity ($\bar{\Cc}_{\mathrm{OFDM}}$) and number of multiplications per unit time ($\Cc_{\mathrm{OFDM}}$):
\begin{equation}
    \label{eq:ofdm_txrx_comp}
    \begin{aligned}
    \bar{\Cc}_{\mathrm{OFDM}}^{\txx/\rxx}\!&=\!\bar{\Cc}_\FFT(M)+4(M\!+\!N_\CP),\\ \bar{\Cc}_{\mathrm{OFDM}}^{\eqlu}\!&=\!4M,\\
    \Cc_{\mathrm{OFDM}}^{\txx/\rxx}\!&=\!\frac{N(\bar{\Cc}_{\mathrm{OFDM}}^{\txx/\rxx}+\bar{\Cc}_{\mathrm{OFDM}}^{\eqlu})}{N(M\!+\!N_\CP)T_\samp}\!=\!\frac{\bar{\Cc}_{\mathrm{OFDM}}^{\txx/\rxx}+\bar{\Cc}_{\mathrm{OFDM}}^{\eqlu}}{M\!+\!N_\CP}F_\samp.
        \end{aligned}
\end{equation}
Furthermore, in the case of DFT-s-OFDM (Fig.~\ref{fig:SCMC_scheme_block_diagram}(b)), the equalization complexity remains the same, while an additional precoding FFT/IFFT block in Tx/Rx results in
\begin{equation}
    \label{eq:dftsofdm_txrx_comp}
    \begin{aligned}
    \bar{\Cc}_{\mathrm{DFTsOFDM}}^{\txx/\rxx}\!&=\!\bar{\Cc}_\FFT(M)+\bar{\Cc}_\FFT(\bar{M})+4(M\!+\!N_\CP),\\
    \Cc_{\mathrm{DFTsOFDM}}^{\txx/\rxx}\!&=\!\frac{\bar{\Cc}_{\mathrm{DFTsOFDM}}^{\txx/\rxx}+\bar{\Cc}_{\mathrm{DFTsOFDM}}^{\eqlu}}{M\!+\!N_\CP}F_\samp.
    \end{aligned}
\end{equation}

SC-FDE enjoys relatively low Tx complexity as symbols are directly transmitted after CP (Fig.~\ref{fig:SCMC_scheme_block_diagram}(c)). However, with FFT/IFFT at Rx, the overall transceiver complexity is that of CP-OFDM (complexity shift from Tx to Rx); $\bar{\Cc}_{\mathrm{OFDM}}\!=\!\bar{\Cc}_{\mathrm{SCFDE}}$.
For OQAM/FBMC, we consider the direct form polyphase prototype filter realization, with a filter length of $L_\mathrm{p}\!=\!O\!\times\!M$ ($O$ is the pulse-shaping overlapping factor).
In general, a multi-tap channel equalization per subcarrier with an equalizer of length $L_{\mathrm{eq}}$ is used for this waveform. Accounting for OQAM, phase offsets (for linear phase filters), IFFT, filtering, $50\%$ overlapping, and equalization, $\bar{\Cc}_{\mathrm{FBMC}}$,  and $\Cc_{\mathrm{FBMC}}$ add up to~\cite{bellanger2010fbmc}
\begin{equation}
    \label{eq:oqamfbmc_txrx_comp}
    \begin{aligned}
    \bar{\Cc}_{\mathrm{FBMC}}^{\txx/\rxx}\!&=\!2\bar{\Cc}_\FFT(M)+4L_\mathrm{p}+4M,\\
    \bar{\Cc}_{\mathrm{FBMC}}^{\eqlu}\!&=\!4ML_{\mathrm{eq}},\\ \Cc_{\mathrm{FBMC}}^{\txx/\rxx}\!&=\!\frac{N(\bar{\Cc}_{\mathrm{FBMC}}^{\txx/\rxx}+\bar{\Cc}_{\mathrm{FBMC}}^{\eqlu})}{M(N\!+\!O\!-\!1/2)}F_\samp,
   \end{aligned}
\end{equation}
where the first multiplication by a factor of $2$ accounts for complex-valued QAM symbols that are separated into two real-valued symbols. The OQAM/FBMC complexity is slightly dependant on ($O$).
Note that we only assume a one-tap equalizer in simulations ($L_{\mathrm{eq}}\!=\!1$). OQAM/FBMC is clearly more complex than CP-OFDM. While for OTFS~\footnote{In this work, we use OTFS with rectangular Tx and Rx windowing and pulse-shaping, and consider one CP per frame ($M\!\times\!N$ symbols), which results in a low-complexity implementation \cite{farhang2017low}. This setting is different from the OTFS setting in~\cite{hadani2017orthogonal} with complexity $\bar{\Cc}_{\mathrm{OTFS}}^{\txx/\rxx}\!=\!2\bar{\Cc}_\FFT(M)+\bar{\Cc}_\FFT(N)+4(N\!+\!N_\CP/M)$ and the OFDM-based OTFS setting in~\cite{surabhi2019otfs} which adds one CP every $N$ blocks (each block is of length $M$). See Sec.~\ref{sec:otfs} for more details.}, based on~\eqref{eq:tx_otfs_vector}, the complexity and number of multiplications per unit time are expressed as
\begin{equation}
    \label{eq:otfs_txrx_comp}
        \begin{aligned}
        \bar{\Cc}_{\mathrm{OTFS}}^{\txx/\rxx}\!&=\!\bar{\Cc}_\FFT(N)+4(N\!+\!N_\CP/M),\\
        \bar{\Cc}_{\mathrm{OTFS}}^{\eqlu}\!&=\!\mathcal{O}(M^3N^3),\\
        \Cc_{\mathrm{OTFS}}^{\txx/\rxx}\!&=\!\frac{M\bar{\Cc}_{\mathrm{OTFS}}^{\txx/\rxx}+\bar{\Cc}_{\mathrm{OTFS}}^{\eqlu}}{N\!M\!+\!N_\CP}F_\samp.
        \end{aligned}
\end{equation}
Hence, $\bar{\Cc}_{\mathrm{OTFS}}/\Cc_{\mathrm{OTFS}}$ are functions of both $N$ and $M$\footnote{In this work, we adopt the classical linear equalizer for OTFS. However, low-complexity iterative and non-iterative solutions can be used. See Table. 2 in~\cite{zhang2022survey} for an extended comparison of complexity costs.}. Moreover, in the case of DFT-s-OTFS (Fig.~\ref{fig:SCMC_scheme_block_diagram}(f)), following the same logic of DFT-s-OFDM:
\begin{equation}
    \label{eq:dftsotfs_txrx_comp}
    \begin{aligned}
    \bar{\Cc}_{\mathrm{DFTsOTFS}}^{\txx/\rxx}\!&=\!\bar{\Cc}_{\mathrm{OTFS}}^{\txx/\rxx}+\bar{\Cc}_\FFT(\bar{N}),\\
    \bar{\Cc}_{\mathrm{DFTsOTFS}}^{\eqlu}\!&=\!\mathcal{O}(M^3N^3),\\
    \Cc_{\mathrm{DFTsOTFS}}^{\txx/\rxx}\!&=\!\frac{M\bar{\Cc}_{\mathrm{DFTsOTFS}}^{\txx/\rxx}+\bar{\Cc}_{\mathrm{DFTsOTFS}}^{\eqlu}}{NM\!+\!N_\CP}F_\samp.
    \end{aligned}
\end{equation}
From \eqref{eq:otfs_txrx_comp} and \eqref{eq:dftsotfs_txrx_comp}, we note that the complexities of OTFS and DFT-s-OTFS are dominated by DD equalization.

\subsection{Peak to Average Power Ratio}
\label{sec:papr_impact}

PAPR is an essential and important KPI for sub-THz/THz communications as it dictates the Tx power efficiency, which affects energy efficiency, link budget, and coverage.
Large amplitude fluctuations in high PAPR lead to spectral regrowth and non-linear distortion; an output back-off is thus needed to retain the linear PA region, reducing power efficiency.
Processing ultra-wide bandwidth sub-THz/THz signals is also very power consuming. Moreover, The saturated output power ($P_\mathrm{sat}$) recordings in state-of-the-art THz PAs \cite{wang2021power} reveal limited achievable output power that decreases drastically with operating frequency (the trend lines for different technologies follow a stepper increasing slope). For example, $P_\mathrm{sat}\!\approx\!\unit[20,23]{dBm}$ and $\unit[28]{dBm}$ at $f_c\!=\!\unit[100]{GHz}$ for CMOS, SiGe BiCMOS, and InP technologies, respectively. Furthermore, high PAPR necessitates high dynamic-range THz ADCs of low signal-to-quantization-noise ratios, which are not cost- and power-efficient \cite{sundstrom2008power}. The ADC signal-to-noise and distortion ratio (SNDR) decreases by increasing the Nyquist sampling rate. However, the energy per conversion step increases linearly with frequencies beyond \unit[100]{MHz} \cite{murmannadc}. For example, for an ADC of $F_\samp\!=\!\unit[100]{GHz}$, the power consumption and SNDR are approximately $\unit[0.3]{Watt}$ (very high) and $\unit[35]{dB}$ (very low), respectively.
The PAPR CCDF of one CP-OFDM symbol $(N\!=\!1)$ is expressed as~\cite{chafii2014closed}
\begin{equation}
    \label{eq:CCDF_theo_ofdm}
   \prob{\mathrm{PAPR}\left(x_{\mathrm{OFDM}}\right)> \gamma_{\mathrm{th}}} = 1- \left(1-e^{-\gamma_{\mathrm{th}}}\right)^M,
\end{equation}
for a PAPR threshold $\gamma_{\mathrm{th}}$. Furthermore, the closed-form approximation of the PAPR CCDF of OQAM/FBMC in \cite{chafii2014closed} reveals higher PAPR values compared to CP-OFDM due to per-subcarrier filtering.
In \cite{surabhi2019peak}, the PAPR CCDF of discrete-time OTFS (no oversampling and rectangular pulse-shaping) is approximated for high values of $N$ as 
\begin{equation}
    \label{eq:CCDF_theo_otfs}
        \prob{\mathrm{PAPR}\left(x_{\mathrm{OTFS}}\right)> \gamma_{\mathrm{th}}}\approx 1- \left(1-e^{-\gamma_{\mathrm{th}}}\right)^{M\!N}.
\end{equation}
The work in~\cite{surabhi2019peak} shows that the PAPR CCDF of OTFS increases with $M$ as the probability of having large peaks increases. However, the maximum OTFS PAPR is upper-bounded by a linear function of $N$ \cite{surabhi2019peak}, unlike TF MC waveforms, such as OFDM, where the PAPR grows linearly with the number of subcarriers $M$.
Note that generalizing \eqref{eq:CCDF_theo_ofdm} over the entire frame approximates \eqref{eq:CCDF_theo_otfs}; OTFS provides significantly better PAPR than OFDM for $N\!<\!M$. Thus, OTFS PAPR is not energy-efficienct for THz system design.
This problem is solved by using a DFT spreading block with OTFS in the uplink~\cite{wu2021dft}, where the PAPR upper bound grows linearly with the DFT spreading size $\bar{N}$.
Since $\bar{N}$ is less than the number of the OTFS symbols in a frame ($N$) and the number of subcarriers ($M$) in a wideband THz channel, $(\bar{N}\!<\!N\!<\!M)$, we expect that DFT-s-OTFS can achieve lower PAPR than both OTFS and OFDM. Thus, DFT-s-OTFS promises to be a more energy-efficient solution for future THz communications. Furthermore, other SCs inherently result in low PAPR, whether in SC-FDE or DFT-s-OFDM, due to DFT-precoding.

\subsection{Robustness to Hardware Impairments}
\label{sec:robust_HW_imp}

THz-band transceivers are substantially more vulnerable to conventional RF impairments than microwave and mmWave transceivers.
Therefore, the waveform's robustness to impairments is a critical KPI. We focus on two important hardware impairments.

\subsubsection{Phase Noise}
\label{sec:phnoise}

Due to time-domain instability, the local oscillator (LO) output can be a phase-modulated tone.
PHN in THz devices (that are not yet mature) has more severe consequences than in microwave or mmWave devices. The motivation to use low-cost devices for THz communications is also limiting, where achieving low PHN requires advanced complex techniques such as phase-locked loops \cite{dahlman20205g}.
In particular, if the THz LO signal is generated using a low-cost low-frequency oscillator followed by frequency multipliers, the required multiplication factor, $\xi$, is relatively high, which further increases the PHN power by a factor of $\xi^2$. Therefore, PHN increases by $\unit[6]{dB}$ for every doubling of the oscillation frequency \cite{dahlman20205g}. Furthermore, PHN causes significant performance degradation and reduces the effective signal-to-interference plus noise ratio (SINR) at the Rx, limiting both data rate and BER. Unfortunately, increasing the signal-to-noise ratio (SNR) does not mitigate the PHN effects. Therefore, optimized SC schemes and non-coherent modulations that are inherently robust to PHN are argued to be good candidates for sub-THz communications \cite{dore2018above}.

There are several approaches for modeling PHN, two of which are most prominent. The first is a correlated model that uses the superposition of Wiener (Gaussian random-walk) and Gaussian processes; the second is an uncorrelated model that considers only a Gaussian noise reflecting the white PHN floor.
The appropriate choice of PHN models for sub-THz band is addressed in \cite{bicais2019phase}, where it is argued that the uncorrelated Gaussian PHN model should be favored if the system bandwidth ($B$) is large enough compared to the oscillator corner frequency ($f_\mathrm{cor}$):
\begin{equation}
    \label{eq:phnmodel_selection_criteria}
    N\left(\frac{f_\mathrm{cor}}{B}\right)^2 \leq \frac{\ln{(2)}}{2\pi}.
\end{equation}
Therefore, the Rx signal, at instant $u$, is expressed as
\begin{equation}
    \label{eq:txrxwithphn_model}
    y[u]=\left(h[u]*\left(x[u]e^{j\phi^{\txx}[u]}\right)\right) e^{j\phi^{\rxx}[u]}+n[u],
\end{equation}
where $*$ denotes linear convolution, and $\phi^{\txx}[u], \phi^{\rxx}[u]$ are discrete stochastic processes representing Tx, Rx LO PHN, respectively.
The correlated model is defined as
\begin{equation}
    \label{eq:ph_corr_model}
    \phi[u]=\phi_{\mathrm{w}}[u]+\phi_{\mathrm{g}}[u],
\end{equation}
where the Wiener and Gaussian PHN models are expressed, respectively, as
\begin{align}
    \label{eq:wiener_corr_model}
    \phi_{\mathrm{w}}[u]=\phi_{\mathrm{w}}[u-1]&+\theta_{\mathrm{w}}[u],\, \theta_{\mathrm{w}}[u]\!\!\sim\!\!\mathcal{N}(0,\sigma^2_\mathrm{w}),\\
    \label{eq:gauss_uncorr_model}
    \phi_{\mathrm{g}}[u]&\!\!\sim\!\!\mathcal{N}(0,\sigma^2_\mathrm{g}).
\end{align}
The uncorrelated PHN implies  $\phi[u]\!=\!\phi_{\mathrm{g}}[u]$. The variances are defined as $\sigma^2_\mathrm{w}\!=\!4\pi^2K_2T$ and $\sigma^2_\mathrm{g}\!=\!K_0/T$, where $K_0$ and $K_2$ are the PHN levels that can be evaluated from the measured PHN PSD, the corner frequency is ($f_\mathrm{cor}\!=\!K_2/K_0$), $T\!=\!1/B$ is the modulated signal duration, and $B$ is the system bandwidth \cite{khanzadi2014calculation}. Thus, we can note a strong dependence of system performance on bandwidth.

\subsubsection{Wideband IQI}
\label{sec:iqimb_comp}

The frequency-dependent wideband IQI is another dominant hardware impairment in THz transceivers operating over ultra-wide bandwidths.
Efficient signal processing techniques have been extensively studied for narrowband IQI at both Tx (via digital pre-distortion) and Rx.
However, only a few works address the wideband IQI model in the THz-band, such as \cite{sha2021channel} for SC-FDE.
Furthermore, wideband PA non-linearity models still lack in the THz literature. Extensive research to study such impairments is crucial. However, the existing models for wideband systems operating at $\unit[60]{GHz}$ \cite{choi2006rf} can provide a preliminary analysis and evaluation of THz candidates.

\subsection{Robustness to THz-specific Impairments}
\label{sec:robust_THz_imp}

THz-specific channel-induced impairments should also be considered when studying candidate waveforms. For example, THz propagation suffers from misalignment between Tx and Rx, which is highly probable given the narrow nature of the THz beams \cite{tarboush2021teramimo}.
Another THz channel characteristic is the spherical wave propagation model (SWM), which should be accounted for at relatively short communication distances \cite{tarboush2021teramimo}. 
More importantly, a beam split effect arises in wideband UM-MIMO beamforming. In particular, the difference between the carrier and center frequencies, $f_m$ and $f_c$, results in THz path components squinting into different spatial directions at different subcarriers, causing severe array gain loss \cite{tarboush2021teramimo}. Such beam split is mainly caused by frequency-independent delays in analog-beamforming PSs. Furthermore, large UM-MIMO THz arrays result in very narrow beamwidths that worsen this effect. Several beam-split mitigation methods are proposed in the literature, such as delay-phase precoding in \cite{dai2021delay}, where CP-OFDM is assumed. However, the effect of beam split on other SC/MC schemes is not yet studied.
This work only studies the impairment caused by beam split as it is more relevant to waveform design than misalignment and SWM.

\section{Candidate THz-band SC/MC Waveforms}
\label{sec:sc_mc_descr}
In the upcoming subsections, we aim to mathematically describe the modulation and demodulation steps for each candidate SC/MC waveform, highlight the design procedure, and link it with THz band system parameters.

\subsection{CP-OFDM}
\label{sec:cpofdm}

The discrete-time Tx OFDM signal is derived from~\eqref{eq:mc_generaltx_dt} ($N\!=\!1$) using rectangular pulse-shaping:
\begin{equation}
    \label{eq:OFDMsig}
    \begin{aligned}
    x_{\mathrm{OFDM}}[u]&=\sum_{m=0}^{M-1}d_\tfd[m]g_\tx[u]e^{j2\pi\frac{m}{M}u},\\
    g_\tx[u] &= 
    \begin{cases}
    \begin{aligned}
            &\frac{1}{\sqrt{M}}&\ u=0,\dots,M\!-\!1\\
            &0&\ \text{otherwise,}
    \end{aligned}
    \end{cases}
    \end{aligned}
\end{equation}
\begin{equation}
    \label{eq:OFDMinmatrixnotation}
    \mbf{x}_{\mathrm{OFDM}} = \mbf{F}^\Hr_M\mbf{d}_M^\tfd. 
\end{equation}
To combat inter-symbol interference (ISI) in a time-dispersive wireless channel of lengh $N_{\mathrm{ch}}\!=\!\tau_\rms/T_\samp$, where $\tau_{\rms}$ is the root mean square (RMS) delay spread, a guard interval of $N_{\CP}\!\geq\! N_{\mathrm{ch}}$ samples is added to the Tx signal.
The CP-OFDM signal can thus be expressed as
\begin{equation}
    \bar{\mbf{x}}_{\mathrm{OFDM}} = \mbf{C}_\CP\mbf{x}_{\mathrm{OFDM}}; \quad \mbf{C}_\CP =
    \begin{bmatrix}
    \mbf{0}_{N_\CP,M\!-\!N_\CP} & \mbf{I}_{N_\CP}\\
    \multicolumn{2}{c}{\mbf{I}_M}
    \end{bmatrix}.
 \label{eq:CPOFDM_matrixnotation}
\end{equation}
where $\bar{\mbf{x}}_{\mathrm{OFDM}}\!=\![x_{\mathrm{OFDM}}[\![-N_\CP]_M\!]\!,\!\cdots\!,x_{\mathrm{OFDM}}[0]\!,\!\cdots\!,\!x_{\mathrm{OFDM}}[M\!-\!1]]^\Tr$, and $\mbf{C}_\CP$ is the CP-insertion matrix of size $M_\trm\!\times\! M$, $M_\trm\!=\!M\!+\!N_\CP$, defined as 
The total CP-OFDM symbol duration, $T\!=\!\left(\!M\!+\!N_\CP\!\right)\!T_\samp\!=\!T_\usf\!+\!T_\CP$, is that of the CP duration ($T_\CP$) plus the useful symbol duration ($T_\usf$).
Although CP reduces SE (Table~\ref{table:SE_comp}), it emulates a cyclic convolution with the channel, allowing simple FDE through FFT.
The received signal over a SISO channel of impulse response $\mbf{h}_{\samp}\!=\![h_0,\dots,h_{N_{\mathrm{ch}}\!-\!1}]^\Tr$, after CP removal is
\begin{equation}
    \mbf{y}_{\mathrm{OFDM}} \!=\!\mbf{H}_{\samp}\mbf{x}_{\mathrm{OFDM}}\!+\!\mbf{n}_{\samp},
    \label{eq:rxsig_cpofdm}
\end{equation}
where assuming perfect time and frequency synchronization, $\mbf{H}_{\samp}$ is an $M\!\times\!M$ circular convolution matrix of band-diagonal structure built upon $\mbf{h}_{\samp}$, and $\mbf{n}_{\samp}\!\!\sim\!\!\mathcal{CN}(\mbf{0},\sigma^2_n\mbf{I}_{M})$ is the AWGN vector. Note that the actual transmission is expressed as $\bar{\mbf{y}}_{\mathrm{OFDM}} \!=\!\bar{\mbf{H}}_{\samp}\bar{\mbf{x}}_{\mathrm{OFDM}}\!+\!\bar{\mbf{n}}_{\samp},$ where $\bar{\mbf{H}}_{\samp}\in\CC^{(M_\trm\!+\!N_{\mathrm{ch}}\!-\!1)\!\times\!M_\trm}$ is derived from $\mbf{H}_{\samp}$, and $\bar{\mbf{n}}_{\samp}\!\!\sim\!\!\mathcal{CN}(\mbf{0},\sigma^2_n\mbf{I}_{M_\trm\!+\!N_{\mathrm{ch}}\!-\!1})$.
The signal is then processed by a DFT block $\mbf{F}_M$. Equalization can be performed using zero-forcing (ZF) or minimum mean-squared error (MMSE), with corresponding equalization matrices
\begin{equation}
    \begin{aligned}
    \mbf{E}_\mathrm{ZF}&= (\mbf{H}_{\samp}^\Hr\mbf{H_{\samp}})^{-1}\mbf{H}_{\samp}^\Hr,\\ \mbf{E}_\mathrm{MMSE}&= (\mbf{H}_{\samp}^\Hr\mbf{H}_{\samp}+\frac{\sigma^2_n}{P_x}\mbf{I}_M)^{-1}\mbf{H}_{\samp}^\Hr,
    \label{eq:equalizer_zfmmse}
    \end{aligned}
\end{equation}
where $P_x$ is the signal power. The Tx symbol estimates are retrieved as $\hat{\mbf{d}}_M^\tfd\!=\!J(\mbf{E}\mbf{F}_M\mbf{y}_{\mathrm{OFDM}})$, where $J(\cdot)$ maps an equalized symbol to the closest symbol in $\mathcal{X}$. Note that \eqref{eq:OFDMinmatrixnotation} can be generalized to express the Tx CP-OFDM frame ($N$ symbols) as
\begin{equation}
    \begin{aligned}
    \bar{\mbf{X}}_{\mathrm{OFDM}}&= \mbf{C}_\CP\mbf{F}^\Hr_M\mbf{D}_\tfd,\\
    \mbf{x}^{\mathrm{OFDM}}_{M_\trm\!N} = \vect{\bar{\mbf{X}}_{\mathrm{OFDM}}}&= \left(\mbf{I}_N\!\otimes\!\mbf{C}_\CP\right)\left(\mbf{I}_N\!\otimes\!\mbf{F}^\Hr_M\right)\mbf{d}_\tfd.
    \label{eq:FrameOFDM_matrix}
    \end{aligned}
\end{equation}

Designing a CP-OFDM system requires tuning many parameters such as the number of subcarriers ($M$), the CP duration ($T_\CP$), and the SCS ($\Delta f$). Such parameters are chosen such that
\begin{equation}
    \tau_\rms\leq T_\CP\leq T_\usf\ll T_\coh; \quad \Delta f=\frac{1}{T_\usf}=\frac{B}{M},
 \label{eq:OFDMdesignEq}
\end{equation}
where $B_\coh\!=\!\frac{1}{5\tau_\rms}$ is the coherence bandwidth and $T_\coh\!=\!\sqrt{\frac{9}{16\pi \nu_\maxx}\times\frac{1}{\nu_\maxx}}$ is the coherence time. The SCS satisfies~\eqref{eq:OFDMdesignEq} to ensure orthogonality and maximize SE.

The SCS choice also affects TTI latency, PAPR, complexity, and equalization performance. In particular, the SCS provides a trade-off between CP overhead, sensitivity to Doppler spread, and robustness to hardware imperfections.
The CP length is also a critical design parameter, where larger $N_{\CP}$ relaxes time synchronization constraints caused by STO, but also at the expense of larger CP overhead (decreased SE). Furthermore, the number of subcarriers ($M$) impacts the PAPR performance \eqref{eq:CCDF_theo_ofdm} and the FFT/IFFT complexity (Sec.~\ref{sec:txrxcomplexity}).

The transmission spectra and molecular absorption dictate the available bandwidth in THz LoS scenarios \cite{tarboush2021teramimo}. However, in indoor THz scenarios, the channel can be LoS-dominant and non-LoS (NLoS)-assisted, or only NLoS (multi-path).
Based on the coherence bandwidth, for a given communication distance, we decide on the corresponding design parameters of a frequency-flat channel or FSC per subcarrier.
For example, for a communication distance of $\unit[3]{m}$, in the sub-THz band ($f_c\!=\!\unit[0.3]{THz}$), $B_\coh\!=\!\unit[1]{GHz}$; in the THz band ($f_c\!=\!\unit[0.9]{THz}$), $B_\coh\!\approx\!\unit[5]{GHz}$ \cite{han2014multi}.

We list in Table~\ref{table:cpofdm_paracomp} some of the expected CP-OFDM parameters for sub-THz/THz band communications, derived using \eqref{eq:OFDMdesignEq} and based on $\tau_\rms$ values from~\cite{han2014multi}, alongside parameters adopted in both 4G-LTE (below $\unit[6]{GHz}$) and 5G-NR (below $\unit[6]{GHz}$ and mmWaves).
In a nutshell, CP-OFDM enjoys a relatively low-complexity implementation (using FFT), is robust to multi-path fading, and uses a simple FDE method (single-tap equalizer for a broadband FSC). However, the resultant high PAPR is challenging for power-limited sub-THz/THz communications.
Moreover, the CP-OFDM time tolerance for symbol synchronization is very low (order of nanoseconds) due to the expected small values of $T_{\CP}$ and $\tau_\rms$.
\begin{table*}[!t]
\footnotesize
\centering
\caption{Key parameters for CP-OFDM at different bands}
\begin{tabular} {|m{4cm}||m{4.2cm}|m{2cm}|m{2cm}|m{2cm}|} 
 \hline \backslashbox{Parameter}{Band} & below $\unit[6]{GHz}$ & mmWave & sub-THz & THz\\
 \hline SCS - $\Delta f(\mathrm{KHz})$ &\begin{tabular}{c}$2^\mu\times15, \mu= 0$ (4G-LTE)\\$2^\mu\times15, \mu= 0, 1, 2$ (5G-NR)\end{tabular} & \begin{tabular}{c}$2^\mu\times15,$\\$\mu= 2, 3, 4$\end{tabular} & \begin{tabular}{c}$2^\mu\times5000,$\\$\mu= 0, 1$\end{tabular} & \begin{tabular}{c}$2^\mu\times25000,$\\$\mu= 0, 1$\end{tabular}\\
 \hline System bandwidth - $B$ $(\mathrm{MHz})$ & \begin{tabular}{c}
 $1.25$ up to $20$ (4G-LTE)\\
 $5$ up to $100$ (5G-NR)
 \end{tabular} & \begin{tabular}{c}$50, 100,$\\$200, 400$\end{tabular}& $[10\!-\!20]\!\times\!10^3$&$[20\!-\!100]\!\times\!10^3$\\
 \hline Number of subcarriers - $M$ & \begin{tabular}{c}
 $128$ up to $2048$ (4G-LTE)\\
 up to $4096$ (5G-NR)
 \end{tabular}  &  up to $4096$ &  up to $2048$&  up to $2048$\\
 \hline Number of used subcarriers &\begin{tabular}{c}
 $76$ up to $1200$ (4G-LTE)\\
 up to $3300$ (5G-NR)
 \end{tabular} & up to $3300$ & up to $1200$& up to $1200$\\
 \hline 
 \end{tabular}
\label{table:cpofdm_paracomp}
\end{table*}
\subsection{DFT-s-OFDM}
\label{sec:dftsofdm}
DFT-s-OFDM, also known as precoded OFDM, is adopted in 4G-LTE/5G-NR uplink and is a promising candidate for THz communications.
The use of a DFT-block at the Tx reduces the PAPR and retains all SC benefits, albeit at a marginal complexity cost.
DFT-s-OFDM thus aims at reducing power consumption and PA costs at user terminal. When data symbol blocks are assigned to different users, DFT-s-OFDM reduces to SC-FDMA in multi-user scenarios.

As illustrated in Fig.~\ref{fig:SCMC_scheme_block_diagram}(b), data symbols are first spread in DFT-precoding; the outputs are the complex symbols that modulate the OFDM subcarriers.
For a selection of $\bar{M}\!\leq\!M$ subcarriers to be modulated. The Tx signal is expressed as
\begin{equation}
        \bar{\mbf{x}}_{\mathrm{DFTsOFDM}} = \mbf{C}_\CP\Hermb{F}_M\left(\mbf{M}_{M,\bar{M}}\mbf{F}_{\bar{M}}\bar{\mbf{d}}_{\bar{M}}^\tfd\right),
 \label{eq:dft-s-ofdm_tx_sig}
\end{equation}
where $\mbf{M}_{M,{\bar{M}}}$ is a mapping matrix between data symbols and the $\bar{M}$ active subcarriers (zero insertion at $M\!-\!{\bar{M}}$ unused subcarriers).
The mapping can be localized or distributed.
In the localized mode, $\mbf{M}_{M,{\bar{M}}} \!=\! [\mbf{I}_{{\bar{M}}},\mbf{0}_{{\bar{M}},M-{\bar{M}}}]^\Tr$, and the DFT outputs are directly mapped to a subset of consecutive subcarriers.
In the distributed mode, the DFT outputs are assigned to non-continuous subcarriers over the entire bandwidth. 
The additional need for signaling, pilots, and guard bands (in multiple access scenarios) in the distributed mode increases the system complexity, whereas the straightforward implementation of equal SCS in the localized mode is favorable.

\subsection{SC-FDE}
\label{sec:scfde}

A promising alternative to CP-OFDM is SC-FDE, which combines the benefits of CP and FDE, and has low PAPR due to low envelope variations.
Unlike in CP-OFDM, where each data symbol is allocated a small bandwidth over a long symbol duration, in SC-FDE, data symbols are assigned to a single large bandwidth with short symbol durations. For the same CP-OFDM symbol duration, the SC-FDE Tx signal, containing $M$ symbols, can be expressed as
\begin{equation}
    \bar{\mbf{x}}_{\mathrm{SCFDE}}=\mbf{C}_\CP\left(\mbf{F}^\Hr_M\mbf{F}_M\right)\mbf{d}_M^\tfd=\mbf{C}_\CP\mbf{d}_M^\tfd.
    \label{eq:scfde_txsig}
\end{equation}
The remainder transmission, equalization, and demodulation stages are similar to those of CP-OFDM, as shown in Fig.~\ref{fig:SCMC_scheme_block_diagram}(c). The SC-FDE synchronization algorithms are also very similar to those of CP-OFDM.
Furthermore, the spectral shape of the SC-FDE waveform is determined by the Tx pulse-shaping, used DAC, and RF filtering stages. It is worth noting that the choice of pulse-shaping affects the PAPR, OOB emissions, complexity, and immunity to hardware impairments.

\subsection{OQAM/FBMC}
\label{sec:oqamfbmc}

OQAM/FBMC is another promising waveform candidate, especially for cognitive radio (CR) and dynamic/intelligent spectrum sharing applications. OQAM/FBMC offers high SE (no need for CP), low OOB emissions levels, and low sensitivity to CFO. Furthermore, by using a per-subcarrier well-localized pulse-shaping filter in both time and frequency (such as PHYDYAS~\cite{viholainen2009deliverable}), OQAM/FBMC supports enhanced synchronization procedures.
However, such benefits come at the cost of limited integration with MIMO systems (maintaining real orthogonality in OQAM complicates precoder design~\cite{perez2016mimo}), higher PAPR compared to OFDM (due to subcarrier filtering), and higher complexity (especially in the equalizer as there is no CP).
Given the importance of such KPIs at high frequencies, OQAM/FBMC is not a good candidate for THz communications.

The direct form of an OQAM/FBMC system is illustrated in Fig.~\ref{fig:SCMC_scheme_block_diagram}(d), consisting of OQAM pre-processing, a synthesis filter bank (SFB), an analysis filter bank (AFB), and OQAM post-processing.
We assume a low-complexity implementation based on a polyphase filter structure (PHYDYAS with overlapping factor $O$) and FFT, as described in \cite{viholainen2009deliverable} (Figures (2-7) and (2-8)).
OQAM/FBMC satisfies the real orthogonality condition, $\mathcal{R}\left(\!\langle\,\!g_{m_1,n_1}^{\mathrm{FBMC}}(t),g_{m_2,n_2}^{\mathrm{FBMC}}(t)\!\rangle\!\right)\!=\!\delta_{(m_2-m_1),(n_2-n_1)}$, instead of complex orthogonality.
Thus, the useful symbol time still satisfies $\Delta f\!=\!1/T_\usf$, but the symbol duration is $T\!=\!T_\usf/2$. The Tx signal can be derived from~\eqref{eq:mc_generaltx} by adding to the Tx basis pulse in~\eqref{eq:mc_generaltx_gt} a phase shift, $\beta_{m,n}=\frac{\pi}{2}(m\!+\!n)$:
\begin{equation}
    g_{m,n}^{\mathrm{FBMC}}(t)=g_\tx^{\mathrm{FBMC}}(t-nT)e^{j2\pi m\Delta f(t-nT)}e^{j\beta_{m,n}}.
    \label{eq:pulseshaping_txfbmc}
\end{equation}
Such a phase shift transfers the induced interference between symbols to the imaginary domain~\cite{nissel2017filter}. The resultant basis pulse in~\eqref{eq:pulseshaping_txfbmc} is a frequency- and time-shifted version of the prototype filter $g_\tx^{\mathrm{FBMC}}(t)$. Furthermore, the prototype filter is designed using the frequency-sampling technique, with ($2O\!-\!1$) non-zero frequency-domain samples for an overlapping factor $O$. For filter of length $L_\mathrm{p}$ and coefficients $\psi[o]$'s (defined in \cite{viholainen2009deliverable}), the impulse response is
\begin{equation}
    g_\tx^{\mathrm{FBMC}}[i] = 1+2\sum_{o=1}^{O-1}(-1)^o\psi[o]\cos\left(\frac{2\pi o}{L_{\mathrm{p}}}(i+1)\right).
    \label{eq:proto_filt}
\end{equation}
Then, the discrete-time Tx signal can be express as
\begin{equation}
    \mbf{x}_{\mathrm{FBMC}}=\mbf{G}_{\mathrm{syn}}\mbf{d}_\tfd,
    \label{eq:fbmcoqam_txmatrix}
\end{equation}
where the time interval is $-OT_\usf/2\!\leq\!t\!<\!OT_\usf/2\!+\!(N\!-\!1)T$ and $\mbf{G}_{\mathrm{syn}}\!\in\!\CC^{N_\trm\times\!M\!N}$ is the Tx matrix that contains the basis pulse-shaping vectors $\mbf{g}_{m,n}^{\mathrm{FBMC}}\!\in\!\CC^{N_\trm\times1}$ ($N_\trm\!=\!\left(OT_\usf\!+\!(N\!-\!1)T\right)F_\samp$), defined as 
\begin{align}
    \label{eq:G_d_fbmcoqam}
    \mbf{G}_{\mathrm{syn}}\!=\![\mbf{g}_{0,0}^{\mathrm{FBMC}}\!\quad\!\cdots\!\quad\!\mbf{g}_{M\!-\!1,0}^{\mathrm{FBMC}}\!\quad\!\mbf{g}_{0,1}^{\mathrm{FBMC}}\!\quad\!\cdots\!\quad\!\mbf{g}_{M\!-\!1,N\!-\!1}^{\mathrm{FBMC}}],&\\
    \mbf{g}_{m,n}^{\mathrm{FBMC}}[n_\trm]=g_{m,n}^{\mathrm{FBMC}}(t)|_{t=n_\trm T_\samp-\frac{OT_\usf}{2}}, n_\trm=0,1,\cdots,N_\trm.
    \label{eq:pulseshaping_vector_fbmc}
\end{align}
At the Rx, the analysis filter, $\mbf{G}_{\mathrm{ana}}\!=\!\mbf{G}^\Hr_{\mathrm{syn}}$, is used in matched-filter decoding.

\subsection{OTFS}
\label{sec:otfs}

The recently proposed OTFS waveform \cite{hadani2017orthogonal} is tailored for high-Doppler doubly-selective channels, typically arising in V2X communications. Unlike the other waveforms that modulate data in the TF domain, OTFS modulates data in the DD domain, transforming the TV channel in TF into a $2$D quasi-TIV channel in DD. The corresponding transmission frame symbols experience a nearly constant channel gain \cite{raviteja2018interference}, making OTFS a promising solution in high-Doppler multi-path channels, exploiting the full diversity of TV-FSC and providing substantial delay and Doppler resilience \cite{surabhi2019diversity}. OTFS is superior to CP-OFDM in this context.

OTFS modulation consists of two main blocks, OTFS transform and Heisenberg transform, as illustrated in Fig.~\ref{fig:SCMC_scheme_block_diagram}(e).
Furthermore, OTFS transform involves two stages, inverse symplectic finite Fourier transform (ISFFT) and windowing. 
ISFFT maps data symbols $d_\ddd[l,k]$ in the DD domain to samples $d_\tfd[m,n]$ in the TF domain as follows \cite{raviteja2018interference}
\begin{equation}
    \label{eq:tx_isfft_otfs}
    d_\tfd[m,n]=\frac{1}{\sqrt{M\!N}}\sum_{l=0}^{M-1}\sum_{k=0}^{N-1}d_\ddd[l,k]e^{j2\pi\left(\frac{nk}{N}-\frac{ml}{M}\right)}.
\end{equation}
A closer look into \eqref{eq:tx_isfft_otfs} reveals that the ISFFT of $\mbf{D}_\ddd$ is equivalent to an $M$-point DFT and an $N$-point IDFT of the columns and rows of $\mbf{D}_\ddd$, respectively.
Subsequently,~\eqref{eq:tx_isfft_otfs} can be expressed in matrix and vectorized forms as
\begin{equation}
    \label{eq:OTFSinmatrixandvector_notation}
    \mbf{D}_\tfd = \mbf{F}_M\mbf{D}_\ddd\mbf{F}^\Hr_N; \quad \mbf{d}_\tfd = (\mbf{F}^\Hr_N\otimes\mbf{F}_M)\mbf{d}_\ddd.
\end{equation}
The OTFS transform applies a Tx window $U_\tx[m,n]$ to the TF signal in \eqref{eq:tx_isfft_otfs}. Let $\mbf{U}_\tx\!=\!\diagg{U_\tx[m,n]}\!\in\!\CC^{M\!N\!\times\!M\!N}$ and assume rectangular windows for both Tx and Rx $\left(\!\mbf{U}_\tx\!=\!\mbf{U}_\rx\!=\!\mbf{I}_{M\!N}\!\right)$, the OTFS transform output is expressed as
\begin{equation}
    \label{eq:tx_otfs_isfftwindow}
    \tilde{\mbf{d}}_\tfd=\mbf{U}_\tx\mbf{d}_\tfd.
\end{equation}
Heisenberg transform then forms the time-domain Tx signal; combining~\eqref{eq:mc_generaltx},~\eqref{eq:mc_generaltx_gt}, and~\eqref{eq:tx_otfs_isfftwindow} 
\begin{equation}
    \label{eq:tx_heisenberg_otfs}
    x_{\mathrm{OTFS}}(t)=\sum_{m=0}^{M-1}\sum_{n=0}^{N-1}\tilde{d}_\tfd[m,n]g_\tx(t-nT)e^{j2\pi m\Delta f(t-nT)}.
\end{equation}
The Tx ($g_\tx(t)$) and Rx ($g_\rx(t)$) pulses ideally satisfy the bi-orthogonality condition \cite{hadani2017orthogonal}, although not practical. Let $\mbf{G}_\tx\!=\!\diagg{g_\tx[0],g_\tx[T\!/\!M],\!\dots\!,g_\tx[(m-1)T\!/\!M]}\!\in\!\CC^{M\!\times\!M}$ be formed from samples of $g_\tx(t)$; $\mbf{G}_\rx$ similarly defined (assuming rectangular pulse-shaping $\mbf{G}_\tx\!=\!\mbf{G}_\rx\!=\!\mbf{I}_M$~\cite{raviteja2018interference}, then $\tilde{\mbf{D}}_\tfd\!=\!\mbf{D}_\tfd$). We can restructure \eqref{eq:tx_heisenberg_otfs} in matrix and vectorized forms as
\begin{align}
    \label{eq:tx_otfs_matrix}
    \mbf{X}_{\mathrm{OTFS}}&=\mbf{G}_\tx\mbf{F}^\Hr_M\left(\mbf{F}_M\mbf{D}_\ddd\mbf{F}^\Hr_N\right)=\mbf{G}_\tx\mbf{D}_\ddd\mbf{F}^\Hr_N,\\
    \label{eq:tx_otfs_vector}
    \mbf{x}_{\mathrm{OTFS}}&=\left(\mbf{F}^\Hr_N\otimes\mbf{G}_\tx\right)\mbf{d}_\ddd.
\end{align}
If $g_\tx(t)$ is a rectangle pulse-shape of duration $T$,~\eqref{eq:tx_heisenberg_otfs} reduces to IDFT, and for $N\!=\!1$, the inner box of Fig.~\ref{fig:SCMC_scheme_block_diagram}(e) is CP-OFDM. Therefore, one OTFS frame is effectively an ISFFT over $N$ consecutive independent OFDM symbols with $M$ subcarriers.

As a spectral-efficient solution, we assume one CP for the entire OTFS frame, of the same duration $T_\CP$ as in previous waveforms.
The Rx signal can be expressed as 
\begin{equation}
    \label{eq:rx_otfs}
    y_{\mathrm{OTFS}}(t)\!=\!\!\!\int_{\!\nu}\!\!\int_{\!\tau}\!\! h_\ddd(\tau,\nu)x_{\mathrm{OTFS}}(t-\tau)e^{j2\pi\nu(t-\tau)}\mathrm{d}\tau \mathrm{d}\nu\!+\!n(t),
\end{equation}
where $\tau$ and $\nu$ are delay and Doppler variables, respectively, and $h_\ddd(\tau,\nu)$ is the DD channel response that is typically sparse ~\cite{raviteja2018interference} (a small number of reflectors with associated delays and Doppler shifts; limited number of multi-paths) and can be expressed as~\cite{raviteja2018interference} 
\begin{equation}
    \label{eq:ch_delayDoppler_otfs}
    h_\ddd(\tau,\nu)=\sum_{i=1}^{N_\mathrm{P}}h_i\bar{\delta}(\tau-\tau_i)\bar{\delta}(\nu-\nu_i),
\end{equation}
where $N_\mathrm{P}$ is the number of paths, $\bar{\delta}(\cdot)$ is the Dirac delta function, and $h_i, \tau_i$, and $\nu_i$ are the $i\thh$-path gain, delay, and Doppler shift, respectively:
\begin{equation}
    \label{eq:delay_Doppler_intOTFS}
    \tau_i=\frac{l_{\tau_i}}{M\Delta f},\,\,\,\,\,\,\nu_i=\frac{k_{\nu_i}}{NT},
\end{equation}
for integers $l_{\tau_i}, k_{\nu_i}$ (indexes of the lattice in \eqref{eq:lattice_dd}). 
Note that the assumptions in \eqref{eq:delay_Doppler_intOTFS} can be further extended to involve fractional Doppler shifts, which result in additional inter-Doppler interference.
The resultant performance degradation can be compensated in the equalizer, using the message-passing algorithm \cite{raviteja2018interference}, for example.
We can ignore fractional delays in a typical wideband THz system since the resolution is sufficient to approximate the path delay to the nearest point in the DD lattice~\cite{raviteja2018interference}. The Rx signal, after discarding CP, is sampled as
\begin{align}
    \label{eq:rx_discrete_nocp_otfs}
    y_{\mathrm{OTFS}}[u]\!\!&=\!\!\!\!\sum_{i=1}^{N_\mathrm{P}}\!\!h_{i}e^{j2\pi\frac{k_{\nu_i}\left(u-l_{\tau_i}\right)}{M\!N}}\!x_{\mathrm{OTFS}}\left[\![u-l_{\tau_i}]_{M\!N}\!\right]\!\!+\!n_{\mathrm{OTFS}}[u],\\
    \label{eq:rx_discrete_nocp_otfs_vector}
    \mbf{y}_{\mathrm{OTFS}}&=\mbf{H}_\ddd\mbf{x}_{\mathrm{OTFS}}+\mbf{n}_{\mathrm{OTFS}},
\end{align}
where $\mbf{y}_{\mathrm{OTFS}}\!\in\!\CC^{M\!N\!\times1}$, $\mbf{n}_{\mathrm{OTFS}}\!\!\sim\!\!\mathcal{CN}(\mbf{0},\sigma^2_n\mbf{I}_{M\!N})$, and $\mbf{H}_\ddd\!\in\!\CC^{M\!N\!\times\!M\!N}$ is the channel matrix
\begin{equation}
    \label{eq:channel_delayDoppler_matrix}
    \mbf{H}_\ddd=\sum_{i=1}^{N_\mathrm{P}}h_{i}\mbf{\Pi}_{l_{\tau_i}}\mbf{\Delta}^{k_{\nu_i}},
\end{equation}
with $\mbf{\Pi}_{l_{\tau_i}}\!\in\!\RR^{M\!N\!\times\!M\!N}$ being the delay matrix, a forward cyclic shifted permutation of $\mbf{\Pi}$ of delay $l_{\tau_i}$ ($\mbf{\Pi}_{l_{\tau_i}\!=1}\!=\!\mbf{\Pi}$ and $\mbf{\Pi}_{l_{\tau_i}\!=0}\!=\!\mbf{I}_{M\!N}$),
\begin{equation}
    \label{eq:permutation_matrix}
    \mbf{\Pi}= \begin{bmatrix}
    0 & \cdots & 0 & 1\\
    1 & \ddots & 0 & 0\\
    \vdots & \ddots & \ddots & \vdots\\
    0 & \ddots & 1 & 0\\
    \end{bmatrix},
\end{equation}
and $\mbf{\Delta}^{k_{\nu_i}}\!\in\!\CC^{M\!N\!\times\!M\!N}$ is the Doppler shift matrix which modulates the Tx signal with a carrier at frequency $k_{\nu_i}$, where $\mbf{\Delta}=\diagg{\omega^0,\omega^1,\cdots,\omega^{M\!N\!-\!1}}$ and $\omega\!=\!e^{\frac{j2\pi}{M\!N}}$.

The Rx signal is then transformed into TF using Wigner transform, which match filters $y_{\mathrm{OTFS}}(t)$ with an Rx pulse shape, $g_\rx(t)$, and samples it at the lattice points defined in \eqref{eq:lattice_tf}.
The Wigner transform is given by
\begin{equation}
    \label{eq:rx_wigner_otfs}
    \begin{aligned}
    y_\tfd[m,n]\!&=\!A_{g_\rx,y_{\mathrm{OTFS}}}(t,f)|_{t=nT,f=m\Delta f},\\ A_{g_\rx,y}(t,f)\!&=\!\!\int\!\! g_\rx^\ast(t'-t)y_{\mathrm{OTFS}}(t)e^{-j2\pi f(t'-t)}dt'.    
    \end{aligned}
\end{equation}
We can express \eqref{eq:rx_wigner_otfs} (similar to \eqref{eq:tx_otfs_matrix}) after building $\mbf{Y}_{\mathrm{OTFS}}\!\in\!\CC^{M\!\times\!N}$ from $\mbf{y}_{\mathrm{OTFS}}$'s in \eqref{eq:rx_discrete_nocp_otfs_vector} as
\begin{equation}
    \label{eq:rx_otfs_matrix}
    \mbf{Y}_\tfd=\mbf{F}_M\mbf{G}_\rx\mbf{Y}_{\mathrm{OTFS}},
\end{equation}
where $\mbf{Y}_\tfd\!\in\!\CC^{M\!\times\!N}$ consists of elements ${y_\tfd[m,n]}$.
The Rx windowing operation is similar to \eqref{eq:tx_otfs_isfftwindow}. Thus, $\tilde{\mbf{y}}_\tfd\!=\!\vect{\tilde{\mbf{Y}}_\tfd}\!=\!\mbf{U}_\rx\mbf{y}_\tfd$ (in our case $\mbf{U}_\rx\!=\!\mbf{I}_{M\!N}$). Then, the TF domain signal, $\tilde{y}_\tfd[m,n]\!=\!y_\tfd[m,n]$, is mapped back to the DD domain using the symplectic finite Fourier transform (SFFT) as
\begin{align}
    \label{eq:rx_sfft_otfs}
    y_\ddd[l,k]&=\frac{1}{\sqrt{M\!N}}\sum_{m=0}^{M-1}\sum_{n=0}^{N-1}y_\tfd[m,n]e^{-j2\pi\left(\frac{nk}{N}-\frac{ml}{M}\right)},\\
     \label{eq:OTFS_rx_matrixnotation}
    \mbf{Y}_\ddd &= \mbf{F}^\Hr_M\mbf{Y}_\tfd\mbf{F}_N=\mbf{F}^\Hr_M\left(\mbf{F}_M\mbf{G}_\rx\mbf{Y}_{\mathrm{OTFS}}\right)\mbf{F}_N,
\end{align}
where $\mbf{Y}_\ddd\!\in\!\CC^{M\!\times\!N}$ is composed of ${y_\ddd[l,k]}$.
We can write \eqref{eq:OTFS_rx_matrixnotation}, after substituting \eqref{eq:tx_otfs_vector} in \eqref{eq:rx_discrete_nocp_otfs_vector}, in a vectorized form
\begin{equation}
    \label{eq:OTFS_rx_vectornotation}
    \begin{aligned}
    \mbf{y}_\ddd &= \vect{\mbf{Y}_\ddd}= \left(\mbf{F}_N\otimes\mbf{G}_\rx\right)\mbf{y}_{\mathrm{OTFS}}\\  &=\!\left(\mbf{F}_N\otimes\mbf{G}_\rx\right)\mbf{H}_\ddd\left(\mbf{F}^\Hr_N\otimes\mbf{G}_\tx\right)\mbf{d}_\ddd\!+\!\left(\mbf{F}_N\otimes\mbf{G}_\rx\right)\mbf{n}_{\mathrm{OTFS}}\\
    &= \mbf{H}_\ddd^{\mathrm{eff}}\mbf{d}_\ddd+\tilde{\mbf{n}}_{\mathrm{OTFS}},
    \end{aligned}
\end{equation}
where $\mbf{H}_\ddd^{\mathrm{eff}}$ denotes the effective channel matrix in the DD domain, and $\tilde{\mbf{n}}_{\mathrm{OTFS}}$ is the modified noise vector.
OTFS equalization and detection can be applied directly on the vectorized form in \eqref{eq:OTFS_rx_vectornotation}, where message passing is shown to be efficient~\cite{raviteja2018interference}. However, we only consider the linear equalizers ZF/MMSE \cite{surabhi2019low} (not the low-complexity version in \cite{surabhi2019low}) for a fair comparison with other candidate waveforms.

Note that $M$ determines the delay resolution and the channel's maximum supported Doppler spread ($\nu_\maxx$) for a given bandwidth ($B$); $N$ dictates Doppler resolution and latency ($T_f\!=\!NT$).
OTFS system design parameters are thus related (we only choose three from $\Delta f, T, M, N$), where $\Delta f\!=\!B/M\!=\!1/T$ is chosen such that
\begin{equation}
    \label{eq:choose_OTFS_deltaf}
    \nu_\maxx<\Delta f <1/\tau_\maxx.
\end{equation}
Using more subcarriers ($M$) results in smaller SCS ($\Delta f$) for a fixed $B$, which in turn results in a longer slot duration ($T$). The OTFS design should thus observe the maximum latency constraints of novel use cases.
Furthermore, the OTFS PAPR, complexity, and decoding delay are proportional to $N$, so lower $N$ values are favored.
However, increasing frame size (large $N$) results in enhanced BER performance \cite{surabhi2019diversity} (higer diversity). Therefore, a careful trade-off between latency, PAPR, complexity, and performance is crucial with OTFS.

Table~\ref{table:fdmax_comp} presents maximum Doppler spread ($\nu_\maxx$) values in different bands, from below $\unit[6]{GHz}$ to THz. The noted severe changes in $\nu_\maxx$ impose many challenges on both waveform design and receiver components (automatic frequency control range and synchronization).
Frequency synchronization in the presence of large CFO (tens/hundreds of KHz) is challenging for CP-OFDM, even for low user mobility, where complex circuits are required at the Rx side. The resilience of OTFS to CFO and Doppler spreads reduces the need for complex wideband automatic frequency control range circuits, which is much needed in THz communications.

\begin{table*}[t!]
\centering
\footnotesize
\caption{Maximum Doppler spread ($\nu_\maxx$) at different terminal speeds and frequency bands}
\begin{tabular} {|c||c|c|c|c|c|c|c|}
 \hline Frequency band & below $\unit[6]{GHz}$ & \multicolumn{2}{|c|}{mmWave} & \multicolumn{2}{|c|}{sub-THz}& \multicolumn{2}{|c|}{THz}\\
 \hline
  \backslashbox{Speed $\upsilon$ ($\mathrm{km/hr}$)}{cen. freq. $f_c$ (Hz)} & \unit[3]{GHz}& \unit[28]{GHz}& \unit[60]{GHz}&\unit[150]{GHz} &\unit[300]{GHz} &\unit[0.8]{THz} &\unit[1.2]{THz} \\ 
 \hline
 5 & 14 & 130 & 278 & 695 & 1390 & 3706 & 5559 \\ 
 \hline
 30 & 83 & 778 & 1668 & 4170 & 8339 & 22238 & 33356\\ 
 \hline
 120 & 334 & 3113 & 6671 & 16678 & 33356 & 88950 & 133426\\ 
 \hline
 300 & 834 & 7783 & 16678 & 41696 & 83391 & 222376 & 333564 \\ 
 \hline
 500 & 1390 & 12972 & 27797 & 69493& 138985 & 370627 & 555940 \\ 
 \hline
 \end{tabular}
\label{table:fdmax_comp}
\end{table*}

\subsection{DFT-s-OTFS}
\label{sec:dft-s-otfs}

DFT-spread-OTFS~\cite{wu2021dft,wu2022dft} is recently proposed for THz ISAC to improve OTFS's PAPR characteristics and enhance the robustness to Doppler effects.
Although this waveform seems to be a promising candidate for many THz applications, a detailed analysis of complexity, SE, TTI latency, and robustness to PHN and THz-specific impairments is still lacking. We conduct this analysis through this work to draw a fair conclusion.

The block diagram of DFT-s-OTFS is illustrated in Fig.~\ref{fig:SCMC_scheme_block_diagram}(f). One DFT-s-OTFS data frame contains the same number of symbols in an OTFS frame $(\!N\!M\!)$. However, for the case of one user in the uplink, only $\bar{N}\!M$ data symbols ($\bar{N}\!\leq\!N$) are first spread using DFT-precoding, similar to DFT-s-OFDM, followed by a DD mapping. The mapping is expressed via $\mbf{M}^\ddd_{N\!M,{\bar{N}\!M}}$ of size $(N\!M)\times(\bar{N}\!M)$, which forms the data frame by concatenating the DFT-spread data into $\bar{N}\!M$ points and zero-padding on the remaining points $(N\!-\!{\bar{N}})\!M$ to form the DD lattice~\footnote{The downlink transmission is detailed in~\cite{wu2021dft}, alongside the uplink scenario where $\bar{K}=N/\bar{N}$ users are multiplexed along the Doppler axis.}. Then, the same operations of OTFS Tx are applied. Thus, the data matrix, $\mbf{D}_\ddd$ of~\eqref{eq:OTFSinmatrixandvector_notation}, is expressed as 
\begin{equation}
\begin{aligned}
        \mbf{D}_\ddd&= \mbf{M}^\ddd_{N\!M,{\bar{N}\!M}}\mbf{F}_{\bar{N}}\mbf{\bar{D}}_{\bar{N}\!M}\\
        \mbf{d}_\ddd &=\vect{\mbf{D}_\ddd}= \mbf{M}^\ddd_{N\!M,{\bar{N}\!M}}\left(\mbf{I}_M\!\otimes\!\mbf{F}_{\bar{N}}\right)\mbf{\bar{d}}_{\bar{N}\!M}.
 \label{eq:dft-s-otfs_dblk}
 \end{aligned}
\end{equation}

We can perform DD domain equalization for delay-Doppler domain signal estimation first, using linear equalizers (MMSE or ZF), and then perform $\bar{N}$-point IDFT to obtain the Tx symbols~\cite{wu2021dft}. Other low-complexity solutions in~\cite{wu2022dft,wu2021dft} do not guarantee a fair comparison with waveforms that use linear equalizer.
As illustrated in Sec.~\ref{sec:papr_impact}, the maximum PAPR of DFT-s-OTFS is limited by $\bar{N}$; an enhanced PAPR performance compared to both OTFS and CP-OFDM \cite{wu2021dft}.
Moreover, due to the potential full TF channel diversity, OTFS and DFT-s-OTFS outperform reference MC schemes.
All that emphasizes the prospects of DFT-s-OTFS in emerging V2X use cases in THz-enabled B5G/6G. However, the price to pay is in increased Rx detection and DD channel estimation complexity, as illustrated in Sec.~\ref{sec:txrxcomplexity}, especially in the presence of fractional Doppler~\cite{wei2021orthogonal}.

\section{Simulation Results and Discussion}
\label{sec:sim_res}

This section presents the results of extensive simulations investigating relevant waveform KPIs under realistic THz conditions.
The default simulation settings are listed in Table~\ref{table:simulationpara_wf} (modifications are declared subsequently and channel parameters are taken form~\cite{tarboush2021teramimo}). We list in Table~\ref{table:PerfEvalPara} a summary of the waveform comparisons under the studied KPIs.

\begin{table}
\footnotesize
\centering
\caption{Simulation parameters}
\begin{tabular} {|c || c|}
 \hline
 Common parameters & Values\\ [0.5ex] 
 \hline
 \hline
 Operating frequency $f_c$ & $\unit[0.3]{THz}$ \\
 System bandwidth $B$ & $\unit[10]{GHz}$\\
 Number of subcarriers $M$ & $128, 256$\\
 Number of MC/OTFS symbols $N$ & $8, 16, 32$\\
 Modulation scheme & $4$-QAM\\
 Overlapping factor (PHYDYAS) $O$ & $4$\\
 DFT-s-OFDM mode & Localized\\
 Oversampling factor $L$ & 4\\
  \hline
  \hline
 Channel parameters & Values\\ [0.5ex] 
 \hline
 \hline
 Molecular absorption coefficient& $\unit[0.0033]{m^{-1}}$\\
 Cluster arrival rate & $\unit[0.13]{nsec^{-1}}$\\
 Ray arrival rate & $\unit[0.37]{nsec^{-1}}$\\
 Cluster decay factor& $\unit[3.12]{nsec}$\\
 Ray decay factor & $\unit[0.91]{nsec}$\\
 Tx and Rx Antenna Elements gains& $\unit[0]{dBi}$\\
 \hline
\end{tabular}
\label{table:simulationpara_wf}
\end{table}

\subsection{Normalized SE and TTI Latency}
\label{sec:se_compare_res}

\begin{figure}[!t]
   \begin{minipage}{0.48\textwidth}
     \centering
     \includegraphics[width=.99\linewidth]{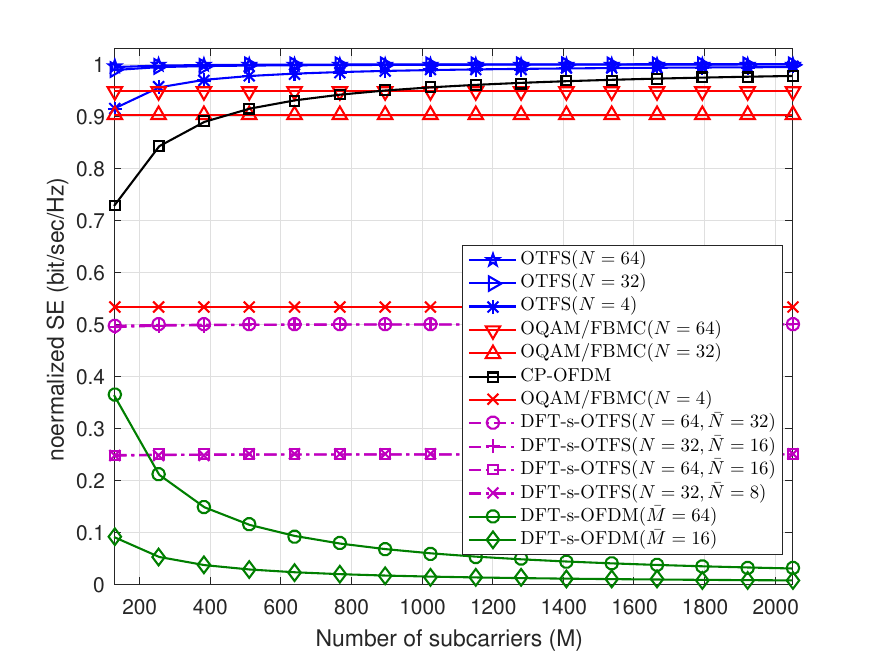}
     \caption{Comparison of normalized SE.}\label{fig:compare_se}
   \end{minipage}\hfill
   \begin{minipage}{0.48\textwidth}
     \centering
     \includegraphics[width=.99\linewidth]{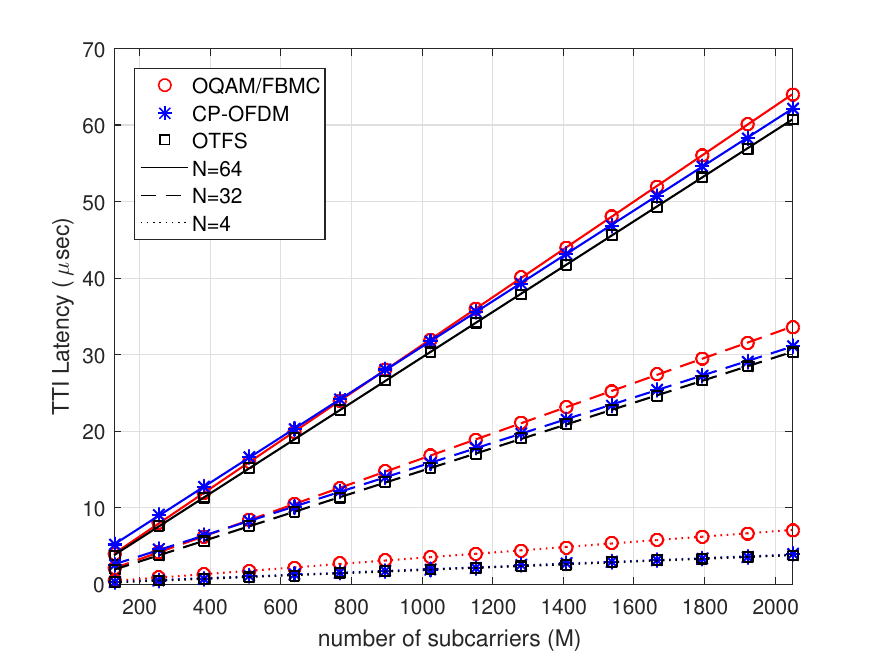}
     \caption{Comparison of TTI latency.}\label{fig:compare_e2e}
   \end{minipage}
\end{figure}

Table~\ref{table:SE_comp} compares the normalized SE and TTI latency of the studied schemes, assuming a fixed modulation order of $\log_2{\left(\abs{\mathcal{X}}\right)}$, and a frame of $M\!N$ symbols; Fig.~\ref{fig:compare_se} further plots the normalized SE values versus the number of subcarriers ($M$).
Assume a target of $\unit[10]{Gbps}$ at a system bandwidth of $B\!=\!\unit[10]{GHz}$, with $N_\CP\!=\!48$. OQAM/FBMC achieves high normalized SE for large $N$ (asymptotic normalized SE of 1 as $N$ goes to infinity; absence of CP) and low normalized SE for short frames (per-subcarrier pulse-shaping extends frame duration by $O\!-\!1/2$).
OTFS achieves the best normalized SE performance, outperforming both CP-OFDM and SC-FDE (mainly due to CP overhead).
An additional SE loss is introduced in DFT-s-OFDM where only $\bar{M}\!\leq\!M$ symbols allocated over $M$ subcarriers \eqref{eq:dft-s-ofdm_tx_sig}.
Moreover, using only $\bar{N}M$ data symbols ($\bar{N}\!\leq\!N$), the DFT-s-OTFS waveform results in a similar SE loss \eqref{eq:dft-s-otfs_dblk}. However, such loss for DFT-s-OFDM is negligible in multi-user scenarios as vacant subcarriers can be allocated to other users, and in DFT-s-OTFS as multi-user can be multiplexed along the Doppler axis (see Fig.\ref{fig:lattice_TF_DD}).
The normalized SE of CP-OFDM increases with $M$, where a CP-OFDM symbol transmits $M$ QAM symbols over $M$ subcarriers of duration $T\!=\!T_\usf+T_\CP$, repeated for $N$ symbols per frame. However, the normalized SE of CP-OFDM does not reach $\unit[1]{bit/sec/Hz}$ and is independent of $N$, which is an important feature for controlling the TTI latency. Note that SC-FDE has the same normalized SE as CP-OFDM; we thus exclude its results (one CP for every $M$ QAM symbols~\eqref{eq:scfde_txsig}).

Regarding TTI latency, Fig.~\ref{fig:compare_e2e} illustrates that OTFS has lower latency than CP-OFDM. DFT-s-OTFS has the same OTFS latency, while SC-FDE and DFT-s-OFDM have the same latency as CP-OFDM.
OQAM/FBMC has lower latency for smaller $N$ and $M$ values, but higher latency for long frame duration.
The OTFS advantages, in terms of normalized SE and TTI latency, are arguably due to the use of a single CP per frame of $M\!N$ symbols, which can be achieved in other waveforms by considering a longer frame of the same number of $M\!N$ symbols and CP length. However, it is important to emphasize that the channel is imposed, and consequently, both $\tau_\rms$ and $T_\coh$ control the maximum symbol and CP duration, as shown for CP-OFDM in~\eqref{eq:OFDMdesignEq}. 
\begin{table}
\footnotesize
\centering
\caption{normalized SE and TTI latency of SC/MC schemes}
\begin{tabular} {|c|c|c|}
 \hline Waveform & normalized SE & TTI latency\\
 \hline CP-OFDM & $\frac{M}{M+N_\CP}$ & $N\!\times\!\frac{M+N_\CP}{F_\samp}$\\
 \hline SC-FDE & $\frac{M}{M+ N_\CP}$& $N\!\times\!\frac{M+N_\CP}{F_\samp}$\\
 \hline DFT-s-OFDM &$\frac{\bar{M}}{M+N_\CP}$& $N\!\times\!\frac{M+N_\CP}{F_\samp}$\\
 \hline OQAM/FBMC & $\frac{N}{N+O-1/2}$ &$M\!\times\!\frac{N+O-1/2}{F_\samp}$\\
 \hline OTFS & $\frac{M}{M+(N_\CP/N)}$ & $\frac{N\!\times\!M+N_\CP}{F_\samp}$\\
 \hline DFT-s-OTFS & $\frac{M\bar{N}}{MN+(N_\CP)}$ & $\frac{N\!\times\!M+N_\CP}{F_\samp}$\\
 \hline
 \end{tabular}
\label{table:SE_comp}
\end{table}

\subsection{PSD and OOB emissions}
\label{sec:psd_res}

The OOB emissions and the impact of adjacent channel leakage are studied in Fig.~\ref{fig:compare_psdwithhole} by comparing the PSD of two users utilizing various waveforms under the IEEE Tx spectral mask specifications (Sec. 13.1.3~\cite{8066476}). Each user occupies a bandwidth of $B\!=\!\unit[2.16]{GHz}$, with $f_c\!=\!\unit[305.64]{GHz}$ for the first user ($\mathrm{ID}\!=\!25$~\cite{8066476}) and the second user is assigned the next channel. The results confirm that OQAM/FBMC has the best frequency localization, thanks to its pulse-shaping filter on each subcarrier; other waveforms respect the specified mask. However, it is expected that a much lower spectral emission mask will be specified for B5G/6G networks. Thus, all waveforms other than OQAM/FBMC would show a high interference level. Note that without additional pulse shaping, the OOB emissions performances of OTFS, DFT-s-OTFS, and SC-FDE are those of CP-OFDM and DFTs-OFDM; thus excluded from Fig.~\ref{fig:compare_psdwithhole}. To ensure a fair comparison with OTFS and DFT-s-OTFS, we concatenate $N\!=\!16$ MC/SC signals in a frame for the previous waveforms.
Although DFT-s-OFDM suffers from high OOB emissions, variants such as zero-tail DFT-s-OFDM \cite{sahin2016flexible} could overcome this limitation.
Furthermore, the performance in the presence of a pulse-shaping filter depends on parameters such as the roll-off factor, oversampling ratio, and used filter length, raised-cosine (RC) for example, (out of this work's scope). Therefore, including a guard band is crucial to achieving the required OOB emissions and interference levels; a careful trade-off between OOB emissions and SE needs to be maintained.

\subsection{Complexity Analysis}
\label{sec:complexity_res}

\begin{figure}[!t]
   \begin{minipage}{0.48\textwidth}
     \centering
     \includegraphics[width=.99\linewidth]{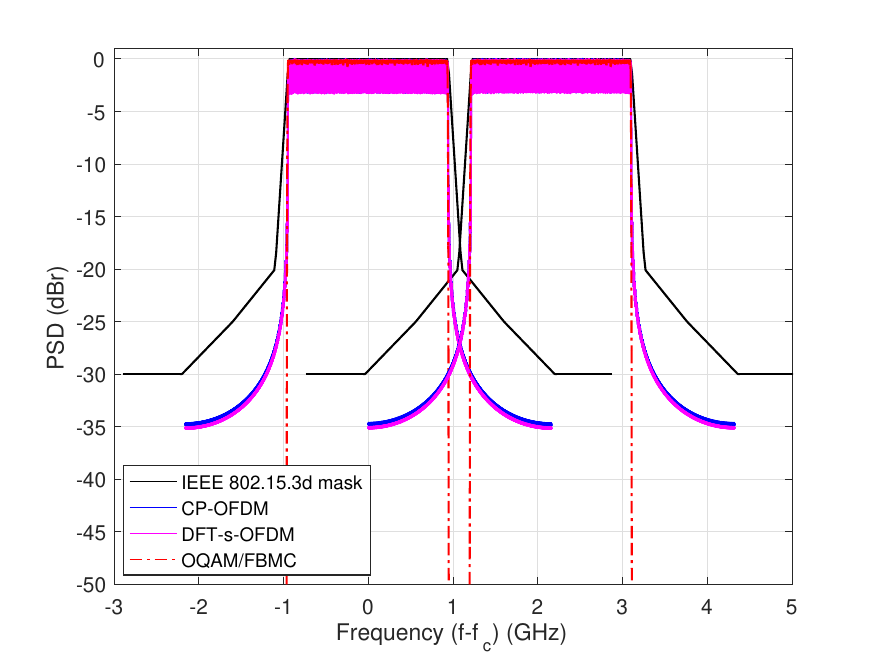}
     \caption{Comparison of OOB emissions between two users' PSD.}\label{fig:compare_psdwithhole}
   \end{minipage}\hfill
   \begin{minipage}{0.48\textwidth}
     \centering
     \includegraphics[width=.99\linewidth]{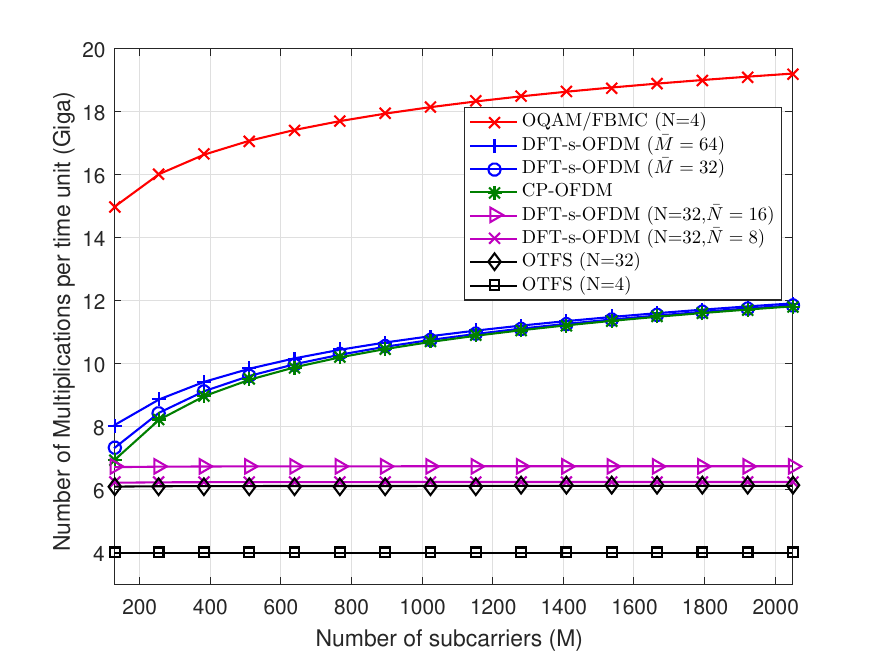}
     \caption{Comparison of Tx/Rx computational complexity without equalization complexity.}\label{fig:compare_complexity}
   \end{minipage}
\end{figure}

A significant part of complexity comes from the equalizer process. It is clear from the analysis in Sec.~\ref{sec:txrxcomplexity} that the DD equalization complexity is multiple orders greater than that of other SC and MC schemes, where the state-of-the-art OTFS equalizers are much more complex than the CP-OFDM ZF/MMSE equalizers. Thus, OTFS and DFT-s-OTFS have a much higher complexity $\mathcal{C}$ than other SC/MC waveforms.
We aim to compare the remaining two components of the complexity (modulation and demodulation).
The computational complexities of different waveforms (Sec.~\ref{sec:txrxcomplexity}) are compared in Fig.~\ref{fig:compare_complexity}, for a maximum $F_\samp\!=\!\unit[1]{GHz}$ (due to hardware constraints) \textbf{but without taking into account the equalization complexity.}
CP-OFDM has the same overall complexity as SC-FDE; CP-OFDM and SC-FDE enjoy lower complexities than both DFT-s-OFDM (due to additional DFT/IDFT precoding blocks in Tx/Rx) and OQAM/FBMC (due to pulse-shaping, overlapping, and OQAM processing that doubles the complexity). OQAM/FBMC is the most complex scheme compared to the previous schemes in that sense.
When neglecting the equalization complexity for both OTFS and DFT-s-OTFS, Fig.~\ref{fig:compare_complexity} shows these two waveforms to be the least complex (complexity that is a function of $N$ and $\bar{N}$; not $M$).
Thus, a viable OTFS solution with low-complexity implementation and good performance, such as unitary approximate message passing (UAMP) or variational Bayes (VB) detection, is a must~\cite{zhang2022survey}.
A low-complexity solution, compared to linear MMSE, is proposed in~\cite{wu2022dft} for DFT-s-OTFS, based on the conjugate gradient method, which has an overall complexity of $\mathcal{O}(M\!N\!\log_2(M\!N))$; the solution still results in high complexity compared to TF MC schemes.

\subsection{CCDF of PAPR}
\label{sec:ccdf_papr_res}
\begin{figure}
  \centering
  \includegraphics[width=0.5\textwidth]{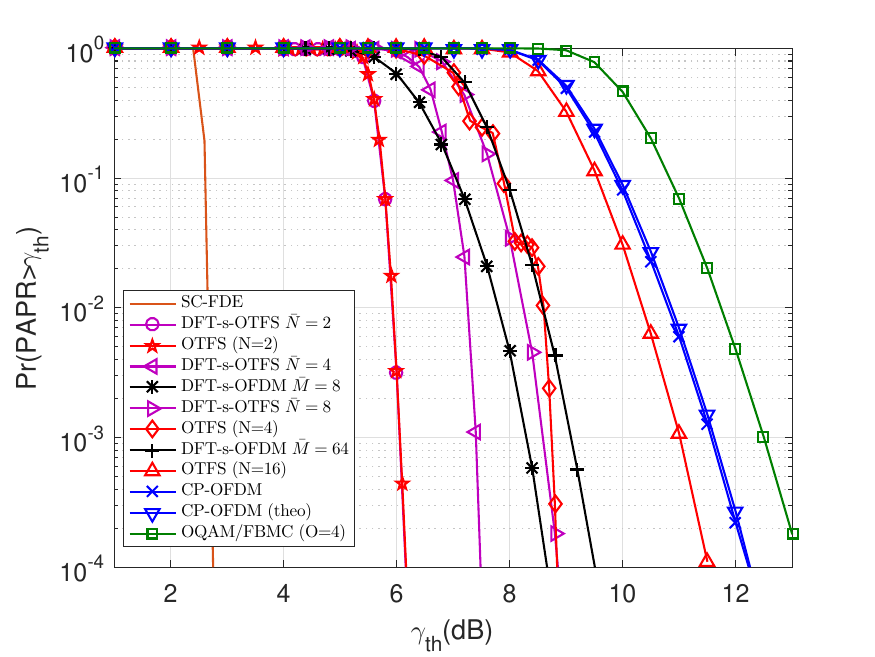}
  \caption{Performance evaluation of PAPR CCDF among waveforms using $16$-QAM, $M\!=\!128$, $N\!=\!16$, rectangular pulse-shaping, and Nyquist sampling.}
  \label{fig:compare_papr}
\end{figure}
\begin{figure*}[t]
  \centering
  \subfloat[CP-OFDM versus OTFS in short frame scenario ($N\!=\!4$).]
  {\label{fig:compare_ofdmotfs_shortframe} \includegraphics[width=0.48\linewidth]{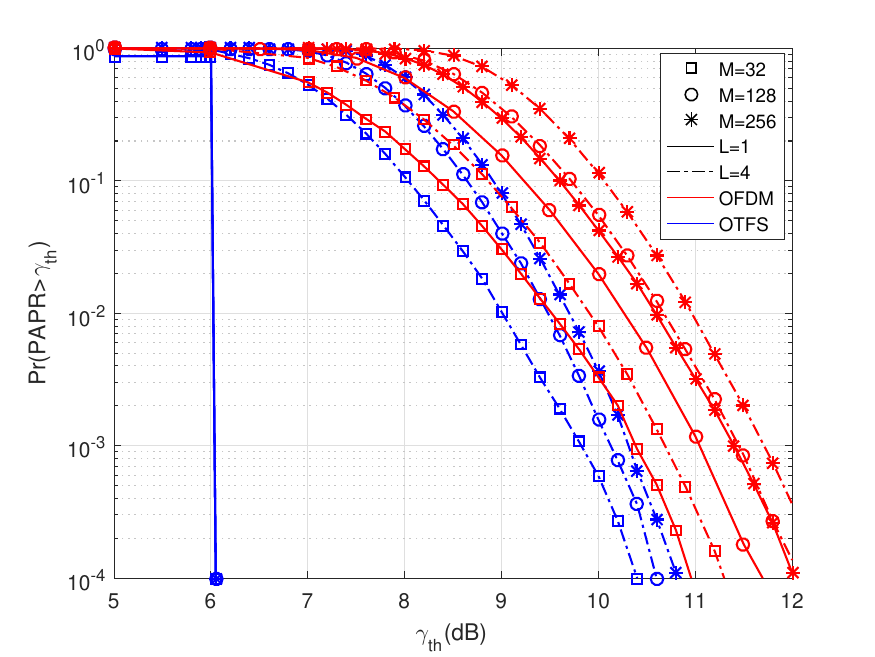}}
  \hfill
   \subfloat[CP-OFDM versus OTFS in long frame scenario ($N\!=\!32$).]
  {\label{fig:compare_ofdmotfs_longframe} \includegraphics[width=0.48\linewidth]{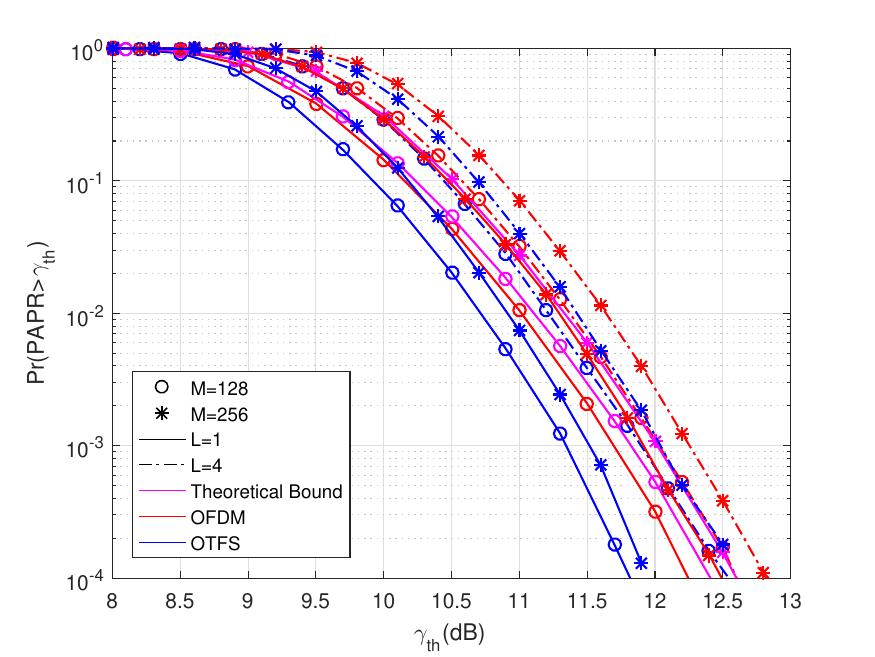}}
  \hfill 
  \subfloat[Effect of pulse-shaping for SC waveforms ($M\!=\!128$ and $N\!=\!16$).]{\label{fig:compare_psf_scfde_dftsofdmdftsotfs} \includegraphics[width=0.48\linewidth]{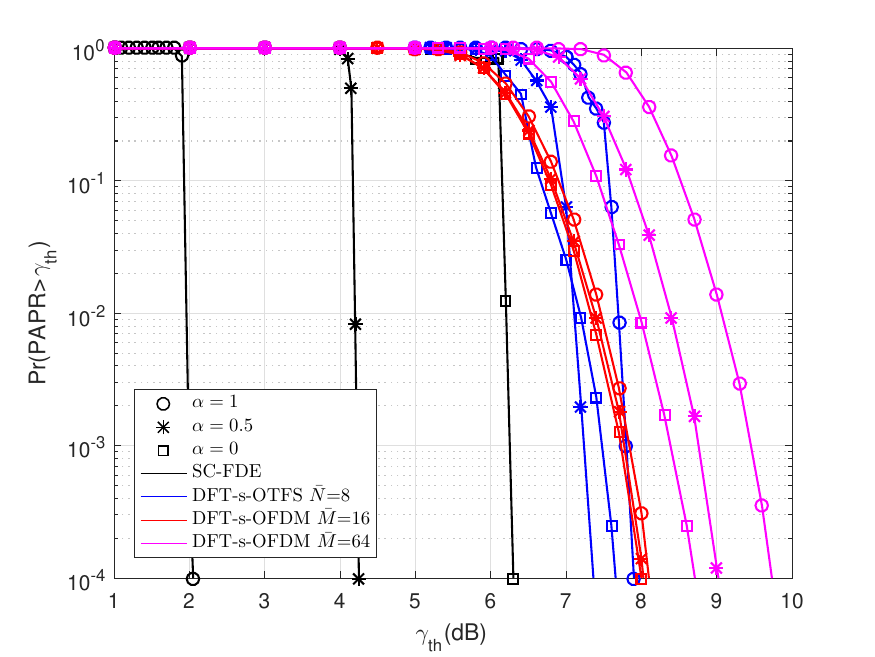}}
  \subfloat[PAPR for discrete and continuous time of DFT-s-OFDM, DFT-s-OTFS, OTFS, and CP-OFDM ($M\!=\!64$ and $N\!=\!32$).]
  {\label{fig:compare_OTFS_OFDM_DFTsOFDM_DFTSOTFS} \includegraphics[width=0.48\linewidth]{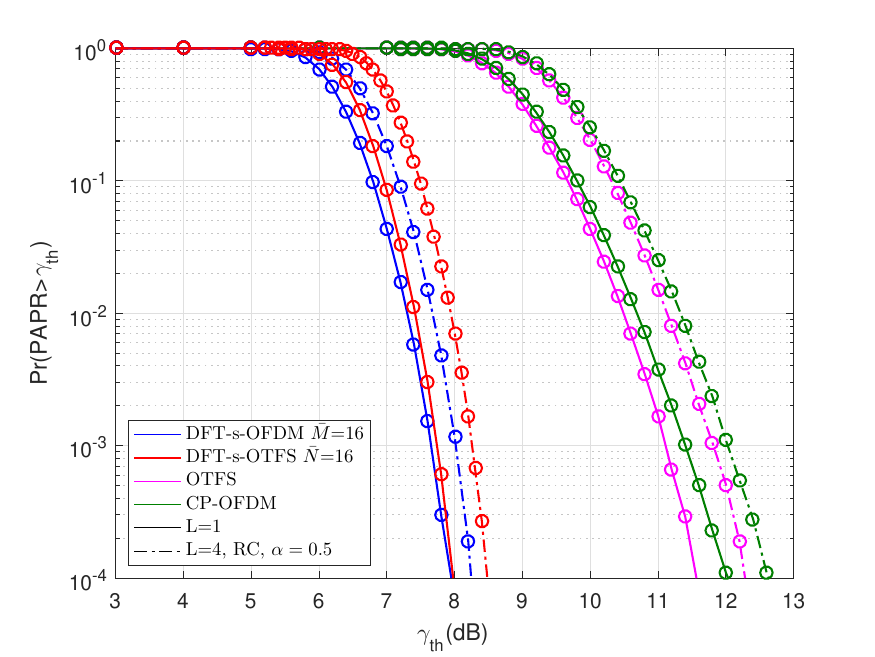}}
  \hfill
  \caption{PAPR CCDF comparison for different scenarios with modulation scheme $4$-QAM.}
  \label{sim_papr}
\end{figure*}
A comparison among the waveforms using different settings and assuming Nyquist sampling is illusrrated in Fig.~\ref{fig:compare_papr}. All schemes (except OQAM/FBMC) apply rectangular pulse-shaping. SC-FDE achieves the best performance (lowest PAPR), followed by DFT-s-OFDM and DFT-s-OTFS, which outperform OTFS, CP-OFDM, and OQAM/FBMC.
In particular, DFT-s-OFDM with $\bar{M}\!=\!8$ and DFT-s-OTFS with $\bar{N}\!=\!8$ show $\unit[3.3]{dB}$ and $\unit[3]{dB}$ PAPR gains compared to CP-OFDM at a CCDF of $10^{-3}$, respectively. Furthermore, DFT-s-OFDM PAPR is dependent on $\bar{M}$ ($\unit[1.8]{dB}$ increase between $\bar{M}\!=\!8$ and $\bar{M}\!=\!64$ at a CCDF of $10^{-3}$) while DFT-s-OTFS PAPR is dependent on $\bar{N}$ ($\unit[1.2]{dB}$ increase between $\bar{N}\!=\!4$ and $\bar{N}\!=\!8$ at a CCDF of $10^{-3}$),
whereas OTFS PAPR is dependent on $N$ \eqref{eq:CCDF_theo_otfs} ($\unit[2.7]{dB}$ increase between $N\!=\!2$ and $N\!=\!4$ at a CCDF of $10^{-3}$).
Note that, as expected from analysis in Sec.\ref{sec:papr_impact}, OTFS shows good characteristics only for small $N$.
OQAM/FBMC is worst performing ($\unit[0.9]{dB}$ worse than CP-OFDM at a CCDF of $10^{-3}$) due to inherent per-subcarrier pulse-shaping.
The CP-OFDM simulations are in agreement with approximation~\eqref{eq:CCDF_theo_otfs}. However, this is not the case for OTFS as the theoretical bound in~\eqref{eq:CCDF_theo_otfs} is only valid for high $N$ values \cite{surabhi2019peak}.

The results of detailed analyses of waveform's PAPR CCDFs are illustrated in Fig. \ref{sim_papr} for different scenarios assuming both Nyquist sampling and oversampling ($L\!=\!4$) to provide accurate conclusions for the discrete and continuous-time signals. Figures~\ref{fig:compare_ofdmotfs_shortframe} and \ref{fig:compare_ofdmotfs_longframe} show the effect of changing $N$ and $M$ on PAPR, with $N\!=\!4, M\!=\!\{32,128,256\}$ and $N\!=\!32, M\!=\!\{128,256\}$, respectively, for both $L\!=\!1\!$ and $L\!=\!4$; we only simulate CP-OFDM and OTFS (CP-OFDM findings also apply to OQAM/FBMC).
We note that increasing either $M$ or $N$ increases PAPR (\eqref{eq:CCDF_theo_ofdm} and~\eqref{eq:CCDF_theo_otfs}).
Nevertheless, the maximum PAPR in OTFS grows linearly with $N$, and the CCDF is zero for $\gamma_{\mathrm{th}}$ values greater than a threshold related to $N$ (for example, the maximum PAPR for $N\!=\!4$ is $10\log{(4)}\!=\!6.02$). For $L\!=\!1\!$, the PAPR gap is more that $\unit[6]{dB}$. However, this gap decreases with continuous-time signals, where the OTFS PAPR gain is no more than $\unit[1.5]{dB}$ for small $N$ values, and is only $\unit[0.3]{dB}$ compared to CP-OFDM at a CCDF of $10^{-3}$ for large $N$ values.
Thus, OTFS provides significantly better PAPR than CP-OFDM only for small $N$ values compared to $M$. However, OTFS still shares the high PAPR characteristics with TF MC signals. Furthermore, we demonstrate in Fig. \ref{fig:compare_ofdmotfs_longframe} good agreement between simulated and analytical results of CP-OFDM for large $N$ and acceptable bound of OTFS (no more than $\unit[0.5]{dB}$ difference).

The effect of pulse-shaping on PAPR performance in SC-FDE, DFT-s-OTFS, and DFT-s-OFDM is illustrated in Fig. \ref{fig:compare_psf_scfde_dftsofdmdftsotfs}, varying the roll-off factor ($\alpha\!=\!\{0, 0.5, 1\}$) of the RC filter: $\frac{\cos{\left(\pi\alpha t/T\right)}\mathrm{sinc}\left(t/T\right)}{\left(1-4\alpha^2(t/T)^2\right)}$ (an RC filter of $6$ symbols; oversampling factor of $4$; normalized to unit energy). We notice that increasing $\alpha$ significantly improves the PAPR performance in SC-FDE, but at the expense of excess bandwidth; DFT-s-OFDM is not as highly affected. Moreover, the PAPR variations with $\alpha$ are negligible for small $\bar{M}$ values. Furthermore, DFT-s-OTFS promises low PAPR when carefully choosing $\bar{N}$ and $\alpha$.

In Fig.~\ref{fig:compare_OTFS_OFDM_DFTsOFDM_DFTSOTFS}, we show that the PAPR in DFT-s-OFDM and DFT-s-OTFS has almost the same value when the DFT precoding sizes are equal ($\bar{N}\!=\!\bar{M}$), for both Nyquist sampling ($L\!=\!1$) and continuous-time signals ($L\!=\!4$). Moreover, the two waveforms secure approximately $\unit[3]{dB}$ PAPR reduction compared with both OTFS and CP-OFDM.

\subsection{Phase Noise}
\label{sec:phn_res}
\begin{figure*}[t]
  \centering
  \subfloat[CP-OFDM performance with three PHN models.]
  {\label{fig:cpofdm_compare_phnmodels} \includegraphics[width=0.48\linewidth]{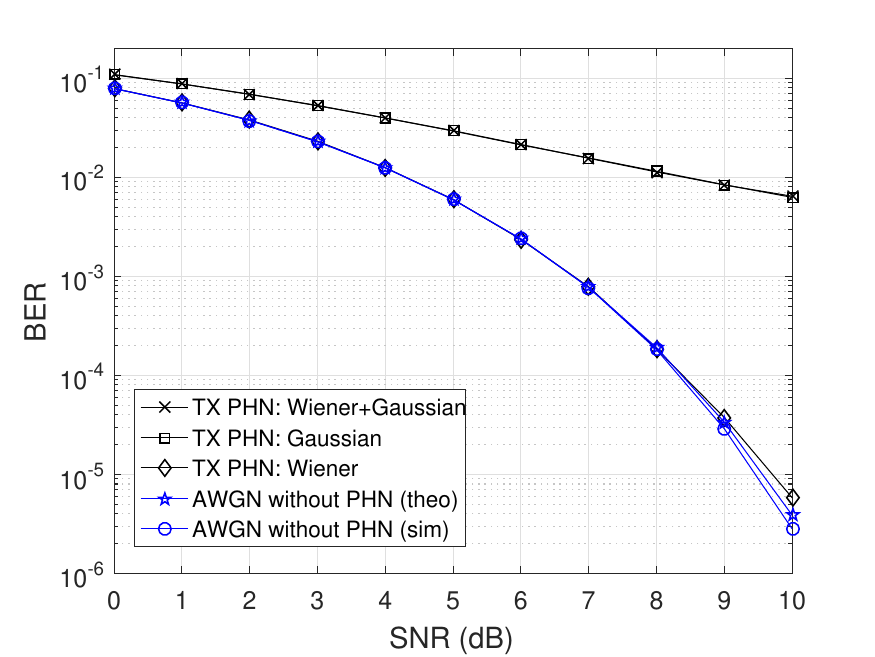}}
  \hfill
  \subfloat[Effect of Gaussian PHN model for different waveforms.]
  {\label{fig:all_compare_gaussphnmodel} \includegraphics[width=0.48\linewidth]{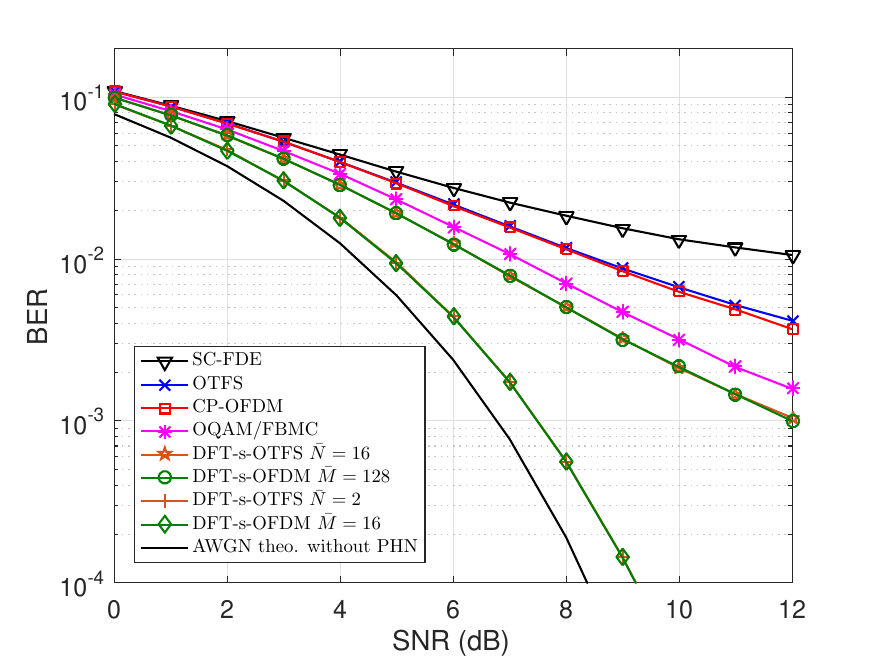}}
  \hfill
  \subfloat[Effect of changing Gaussian PHN variance at $\unit{10dB}$ SNR.]
  {\label{fig:all_changesigmaG_gaussphnmodel} \includegraphics[width=0.48\linewidth]{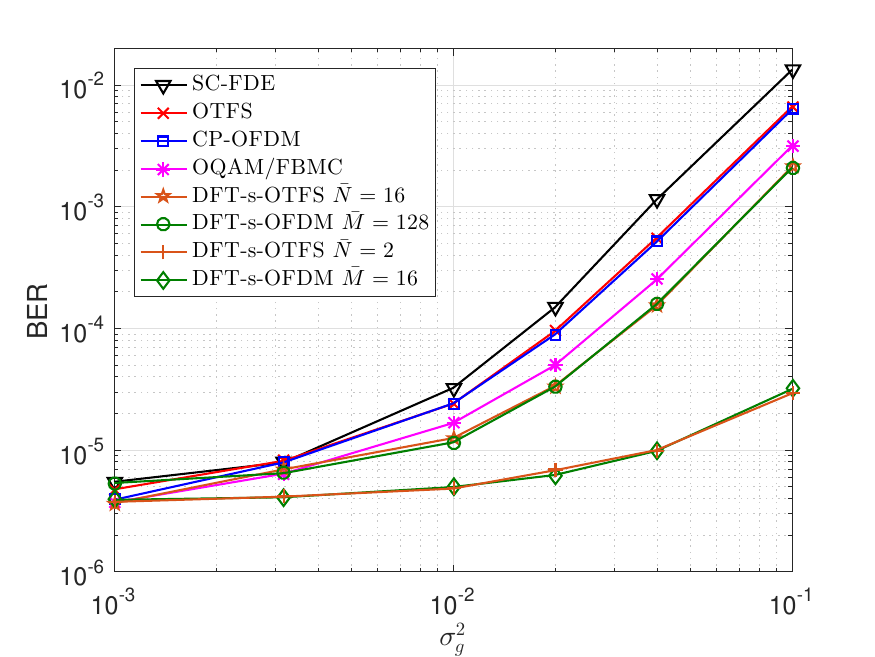}}
  \subfloat[Effect of changing the SCS with Gaussian PHN.]
  {\label{fig:all_changeSCS_gaussphnmodel} \includegraphics[width=0.48\linewidth]{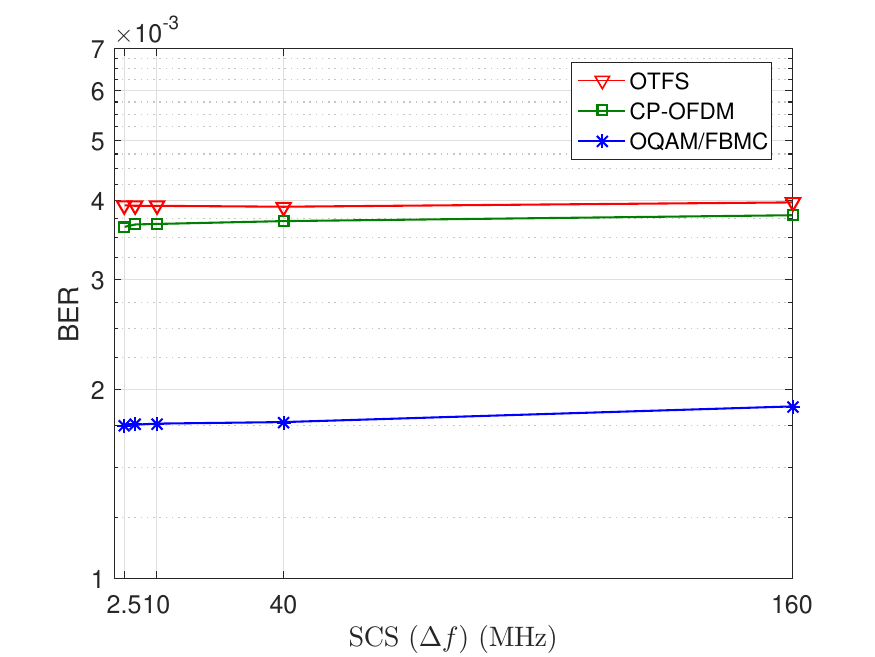}}
  \hfill
  \caption{BER performance of various schemes with Tx PHN ($M\!=\!256$ and $N\!=\!32$).}
  \label{sim_phn}
\end{figure*}
We first verify our assumption of Gaussian PHN for THz communications. We incorporate the PHN measurement results in~\cite{yi2020300} for a $\unit[300]{GHz}$ signal source, a PHN floor level of $K_0\!=\!\unit[-110]{dBc/Hz}$, and $K_2\!=\!10$ ($f_\mathrm{cor}\!=\!\unit[1]{MHz}$); we set $B\!=\!\unit[10]{GHz}$, which satisfies \eqref{eq:phnmodel_selection_criteria}.
We consider Tx PHN without loss of generality. For an AWGN channel plus Tx PHN ($h[u]\!=\!1$, for all $u$ in~\eqref{eq:txrxwithphn_model}), the CP-OFDM results in Fig.~\ref{fig:cpofdm_compare_phnmodels} illustrate that the models in~\eqref{eq:ph_corr_model} and~\eqref{eq:gauss_uncorr_model} are equivalent. Hence, for large bandwidths, the uncorrelated model of~\eqref{eq:gauss_uncorr_model} is sufficient.
The reason behind this observation is that the Wiener model PSD decreases with frequency, resulting in PHN power levels lower than the white floor noise of the Gaussian model at frequencies higher than ($f_\mathrm{cor}$). Note that the PHN power of the Gaussian model is constant; we add a reference AWGN lower bound.
PHN leads to inter-carrier interference (ICI) and adjacent channel interference, which explains the resultant degradation.
The comparison assuming Gaussian PHN in Fig.~\ref{fig:all_compare_gaussphnmodel} illustrates that DFT-s-OFDM and DFT-s-OTFS are the most robust waveforms to PHN, and decreasing $\bar{M}$ and $\bar{N}$ enhances the performance. 
Surprisingly, we demonstrate that both DFT-s-OFDM and DFT-s-OTFS result in the same performance when the ratios $M/\bar{M}$ and $N/\bar{N}$ are equal.
OQAM/FBMC outperforms other schemes because of its good time and frequency localization. Furthermore, CP-OFDM and OTFS are more robust than SC-FDE. SC-FDE has the worst performance.
We also analyze the effect of changing the noise variance ($\sigma^2_\mathrm{g}$) assuming THz-band Gaussian PHN and an SNR of $\unit[10]{dB}$ in Fig.~\ref{fig:all_changesigmaG_gaussphnmodel}. We vary $\sigma^2_\mathrm{g}$ between low ($10^{-3})$, medium ($10^{-2}$), and strong ($10^{-1}$) values (as indicted in Table I in~\cite{bicais2019phase}), retaining a system bandwidth of $B\!=\!\unit[10]{GHz}$; this changes the spectral density ($K_0$) of the white PHN floor. 
Increasing $\sigma^2_\mathrm{g}$ increases the BER, where DFT-s-OFDM and DFT-s-OTFS are the best performing. 

In Fig.~\ref{fig:all_changeSCS_gaussphnmodel}, we study the effect of changing SCS by changing $M\!=\!\{4096,2048,1024,256,64\}$ and fixing $B\!=\!\unit[10.24]{GHz}$. Surprisingly, the waveform BERs are retained, which is an important feature that relaxes other design parameters. For example, we can use a small $M$ to ensure low PAPR and high PHN robustness concurrently. Such results are not observed below $\unit[6]{GHz}$, where increasing SCS ensures high robustness to PHN (different low-frequency models).

\subsection{Beam Split}
\label{sec:beamsplit_res}

\begin{figure}[!t]
   \begin{minipage}{0.48\textwidth}
     \centering
     \includegraphics[width=.99\linewidth]{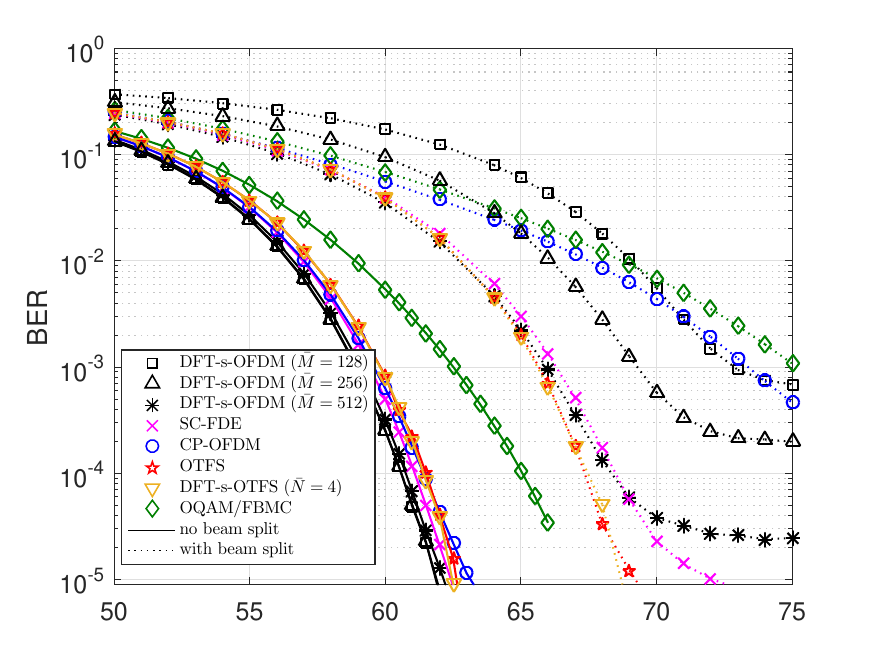}
     \caption{BER performance of various waveforms in the presence and the absence of beam split ($M\!=\!1024$ and $N\!=\!8$).}\label{fig:compare_beamsplit}
   \end{minipage}\hfill
   \begin{minipage}{0.48\textwidth}
     \centering
     \includegraphics[width=.99\linewidth]{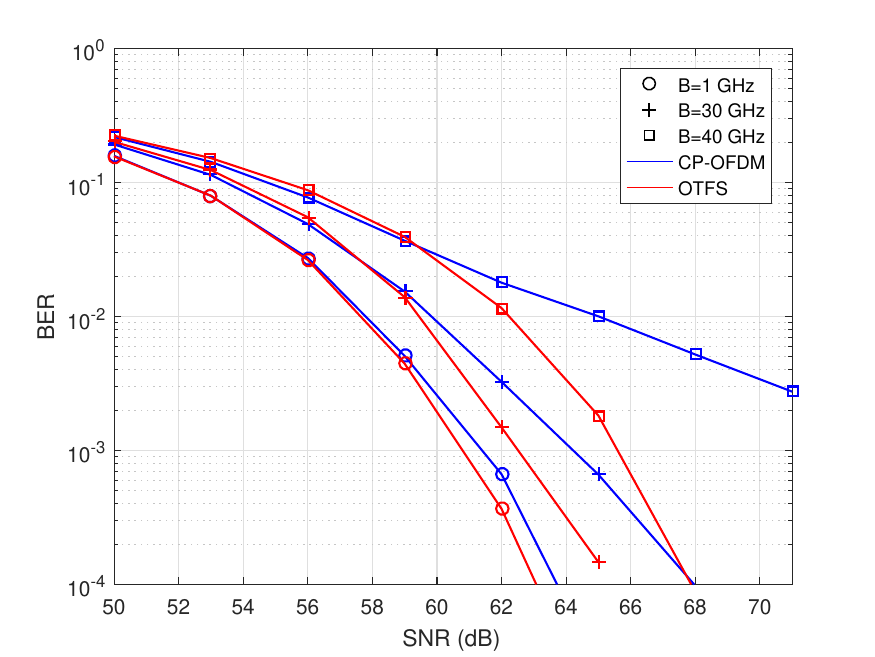}
     \caption{Effect of changing the systen bandwidth $B$ in the presence of beam split ($M\!=\!512$ and $N\!=\!4$).}\label{fig:chnage_BW_BSE}
   \end{minipage}
\end{figure}

We compare all waveforms using a stochastic THz channel simulator, TeraMIMO \cite{tarboush2021teramimo}, in Fig.~\ref{fig:compare_beamsplit}, for $B\!=\!\unit[50]{GHz}$ and $f_c\!=\!\unit[0.325]{THz}$.
We consider an UM-MIMO system with beamforming, where both Tx and Rx have uniform linear arrays of $32$ AEs; we set the communication distance to $\unit[1]{m}$. The channel is LoS-dominant with a few multi-path components, which tends to be almost flat-fading.
We first plot the BERs assuming the absence of beam split; all waveforms achieve similar performance except for OQAM/FBMC (we considered simple MMSE equalization). When adding beam split, OQAM/FBMC is shown to be less affected compared to CP-OFDM. DFT-s-OTFS, OTFS, SC-FDE, and DFT-s-OFDM (with large $\bar{M}$) have high robustness to THz-induced impairments, securing multiple-dB BER gains over both CP-OFDM and OQAM/FBMC. Furthermore, we study the effect of changing the system bandwidth $B$ for both CP-OFDM and OTFS. We keep the previous simulation settings and only change the system bandwidth as $B\!\in\!\{1,30,40\} \mathrm{GHz}$. Fig.~\ref{fig:chnage_BW_BSE} shows that increasing the system bandwidth $B$ results in severe performance degradation due to significant array gain loss. The THz path components squint into different spatial directions at different subcarriers, causing this loss. Moreover, the results confirm the superiority of OTFS compared to CP-OFDM in terms of beam split robustness.

\subsection{Performance in Doubly-Selective Channels}
\label{sec:tvdoppler_res}

\begin{figure}
\centering
\includegraphics[width=0.5\textwidth]{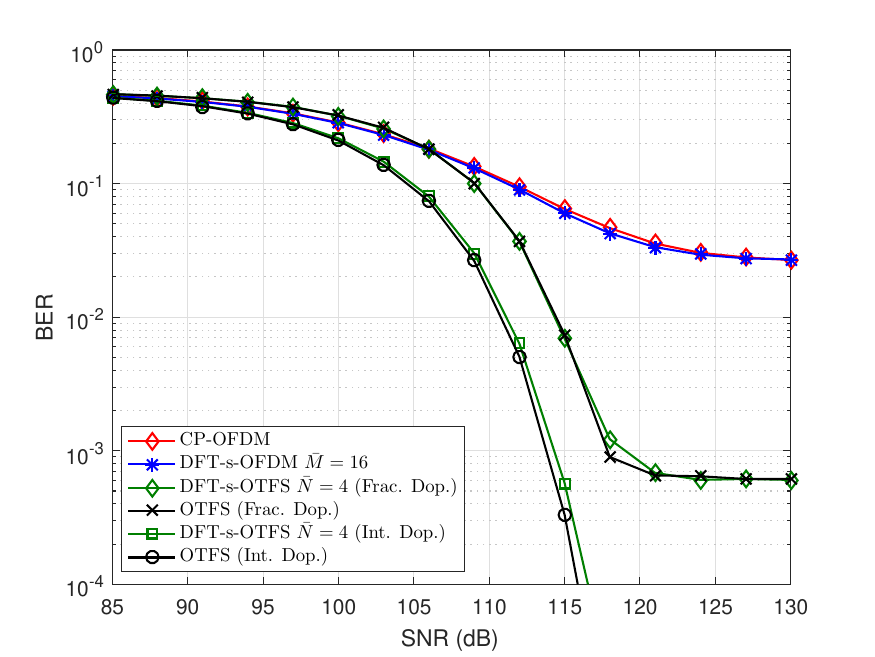}
\caption{BER performance of various schemes in a THz TV-FSC.} 
\label{fig:compare_tvfsc}
\vspace{-5mm}
\end{figure}

Doppler spreads in THz channels are orders-of-magnitude larger than those in the conventional microwave and mmWave channels (Table~\ref{table:fdmax_comp}).
In Fig.~\ref{fig:compare_tvfsc}, we compare the BERs of the studied schemes in a doubly-selective THz channel. We consider $f_c\!=\!\unit[0.5]{THz}$, $B\!=\!\unit[0.25]{GHz}$, $M\!=\!64$, $N\!=\!16$ (for fairness between DFT-s-OTFS, OTFS and other waveforms, we concatenate $N$ symbols per frame), and user velocity $\upsilon\!=\!\{\unit[500]{km/hr}\}$. Note that the communication distance is $\unit[2]{m}$, and cluster/rays parameters are taken from Table~\ref{table:simulationpara_wf} (waveform parameters are derived following \eqref{eq:choose_OTFS_deltaf}).
We consider MMSE equalization for all waveforms (other waveforms such as OQAM/FBMC show the same performance as CP-OFDM and are thus omitted).
DFT-s-OTFS and OTFS more robust than CP-OFDM and other waveforms in TV-FSC, even for larger user velocity ($\upsilon$), showing multiple-dB BER gains. Such advantages render DFT-s-OTFS and OTFS exceptionally suitable for high-mobility, high-carrier scenarios (THz V2X scenarios, for example).
Note that we consider both integer and fractional Doppler shifts when simulating OTFS and DFT-s-OTFS. We notice that fractional Doppler causes performance degradation due to the inter-Doppler interference. However, OTFS and DFT-s-OTFS are still superior to other SC/MC waveforms in high-speed scenarios.

\subsection{Recommendations}
\label{sec:discuss}

Table~\ref{table:PerfEvalPara} presents a summary of the waveforms' performance under the adopted KPIs in this work. However, the importance of these KPIs varies from one application/use case to another in 6G networks.
We first list some of the 6G use cases mentioned in~\cite{zhang20196g} and match them to the appropriate KPIs.
Typical applications in further-enhanced mobile broadband (FeMBB) scenarios are holographic MIMO, AR, and VR. The relevant KPIs are, but not limited to, enhanced SE, UM-MIMO compatibility, and robustness to beam split (as it is related to the usage of wide bandwidths). Thus, we recommend DFT-s-OFDM as a viable solution, as illustrated in Table~\ref{table:PerfEvalPara}.

Ultra-massive machine-type communications (UMMTC) use cases includes several applications such as the Internet of everything and smart home and city. The major KPIs affecting UMMTC performance are low latency, robustness to hardware impairments (like PHN), and increased energy efficiency (lower PAPR and high SE lead to enhanced energy efficiency). Therefore, DFT-s-OFDM waveforms should be a priority based on our evaluation. Furthermore, for extremely low-power communications (ELPC) use cases, such as the Internet of bio-nano-things, the DFT-s-OTFS seems to be the most promising candidate as energy efficiency is the important KPI.
Extremely reliable and low-latency communications (ERLLC) scenarios involve fully automated driving and industrial Internet. The robustness to doubly selective channels, immunity to high Doppler spreads, and low latency are the determinant KPIs. Thus, we recommend DFT-s-OTFS and OTFS.

Another use case is the THz ISAC, where an energy-efficient waveform with high robustness to Doppler shifts and PHN is desired. We recommend DFT-s-OFDM and DFT-s-OTFS for THz ISAC.
Other 6G verticals impose novel/specific requirements on localization. However, our analysis did not include any KPIs directly related to localization performance.

\section{Conclusion}
\label{sec:conclusion_future_pros}
\begin{table*}[t!]
\footnotesize
\centering
\caption{Performance evaluation metrics for different SC/MC waveforms}
\begin{tabular} {|c||c|c|c|c|c|c|}
 \hline
 \backslashbox{Metric}{Waveform} & SC-FDE & DFT-s-OFDM & CP-OFDM & OQAM/FBMC & OTFS & DFT-s-OTFS\\ [0.5ex]
 \hline
 Spectral efficiency & medium & medium & medium & low/high & high & high\\
 \hline
 TTI latency & low & medium & low/medium & high & low/medium & low/medium\\
 \hline
 UM-MIMO compatibility &high& high& high& low& high& high\\
 \hline
 PAPR &low & low/medium &high & high & medium/high & low/medium\\
 \hline
 OOB emissions & medium/high & high & high & low & high& high\\
 \hline
 Gaussian PHN robustness & low & high & low/medium & medium &low/medium & high\\
 \hline
 Beam split robustness & high & high & low & low/medium & high & high\\
 \hline
 Robustness to doubly selec. chan. & low& low& low& low&high&high\\
 \hline
 Modulator/Demodulator Complexity &medium &medium &medium &high &low/medium &low/medium\\
 \hline
 Equalization Complexity &low &low &low &medium &medium/high &high\\
 \hline
\end{tabular}
\label{table:PerfEvalPara}
\vspace{-3mm}
\end{table*}

In this paper, a comprehensive study of SC/MC waveforms for THz communications is conducted. The analysis and simulation results demonstrate that the candidate 5G waveforms (filtered-based OFDM, such as OQAM/FBMC) are not suitable for future B5G/6G networks because of the increased PAPR and complexity. Furthermore, CP-OFDM and SC-FDE share similar characteristics: good SE, moderate TTI latency, high UM-MIMO compatibility, acceptable to high OOB emissions (without any additional pulse-shaping), and relatively low implementation complexity (especially with a single-tap ZF/MMSE equalizer). SC-FDE is shown to be less robust to uncorrelated Gaussian PHN, but it results in low PAPR and high robustness to THz beam split. DFT-s-OFDM is further shown to offer low PAPR and high robustness to both THz PHN and beam split. Finally, DFT-s-OTFS is illustrated to achieve high SE, low TTI latency, good PAPR characteristics, and high robustness to THz impairments. However, these advantages come at the price of increased equalization complexity, which opens important future research directions. Furthermore, DFT-s-OTFS and OTFS outperform all other waveforms in doubly-selective channels. In a nutshell, the findings of this work recommend the use of DFT-s-OFDM and DFT-s-OTFS in B5G/6G sub-THz/THz communications; CP-OFDM can still be used in sub-THz indoor scenarios (TIV-FSC). Other relevant performance metrics can be considered in future works. For instance, researchers should study the waveform robustness to asynchronous access, synchronization procedures in the presence of both STO and CFO, wideband IQI, PA non-linear distortion, multi-user scheduling, and flexible resource allocation.
\begin{table}
\footnotesize
\centering
\caption{Summary of Frequently-Used Acronyms}
\begin{tabular} {|c|c|} 
 \hline Abbreviation & Definition\\ [0.5ex] 
 \hline 4G-LTE & fourth-generation long term evolution\\ 
 \hline 5G-NR & fifth-generation new-radio\\
 \hline 6G & sixth-generation\\
 \hline ADC & analog-to-digital converter\\
 \hline AE & antenna element\\
 \hline AFB & analysis filter bank\\
 \hline AoSAs & array-of-subarrays\\
 \hline AWGN & additive white Gaussian noise\\
 \hline B5G & beyond-fifth generation\\
 \hline BDMA & beam-division multiple-access\\
 \hline BER & bit error rate\\
 \hline CCDF & complementary cumulative distribution function\\
 \hline CFO & carrier-frequency offset\\
 \hline CP & cyclic-prefix\\
 \hline CPM & continuous phase modulation\\
 \hline CR & cognitive radio\\
 \hline DAC & digital-to-analog converter\\
 \hline DD & delay-Doppler\\
 \hline DFT-s-OFDM & discrete-Fourier-transform spread OFDM\\
 \hline DFT-s-OTFS & discrete-Fourier-transform spread OTFS\\
 \hline ELPC & extremely low-power communications \\
 \hline ERLLC & extremely reliable and low-latency communications\\
 \hline f-OFDM & filtered-OFDM\\
 \hline FBMC & filter-bank multi-carrier\\
 \hline FDE & frequency domain equalization \\
 \hline FDMA & frequency-division multiple-access\\
 \hline FeMBB & further-enhanced mobile broadband\\
 \hline FFT & fast Fourier transform\\
 \hline FSC & frequency-selective channel\\
 \hline GFDM & generalized frequency-division multiplexing\\
 \hline ICI & inter-carrier-interference\\
 \hline IFFT & inverse FFT\\
 \hline IQI & in/quadrature-phase imbalance\\
 \hline ISAC & integrated sensing and communication\\
 \hline ISFFT & inverse SFFT\\
 \hline ISI & inter-symbol-interference\\
 \hline ITU & international telecommunications union\\
 \hline IRS & intelligent reflecting surface\\
 \hline KPI & key performance indicator\\
 \hline LAN & local area network\\
 \hline LoS & line-of-sight\\
 \hline MC & multi-carrier\\
 \hline MHz & megahertz\\
 \hline MMSE & minimum mean-squared error\\
 \hline mmWave & millimeter-wave \\
 \hline NLoS & non-LoS\\
 \hline NOMA & non-orthogonal multiple-access\\
 \hline
\end{tabular}
\label{table:AbbreviationsTable1}
\vspace{-5mm}
\end{table}
\begin{table}
\footnotesize
\centering
\caption{Summary of Frequently-Used Acronyms}
\begin{tabular} {|c|c|} 
 \hline Abbreviation & Definition\\ [0.5ex] 
 \hline OFDM & orthogonal frequency-division multiplexing\\
 \hline OOB & out-of-band\\
 \hline OOK & on-off keying\\
 \hline OQAM & offset quadrature amplitude modulation\\
 \hline OTFS & orthogonal time-frequency space\\
 \hline PA & power amplifier\\
 \hline PAPR & peak-to-average power ratio\\
 \hline PHN & phase noise\\
 \hline PS & phase-shifter\\
 \hline RF & radio frequency\\
 \hline SA & subarray\\
 \hline SC-FDE & single-carrier frequency-domain equalization\\
 \hline SCS & subcarrier spacing\\
 \hline SE & spectral efficiency\\
 \hline SFB & synthesis filter bank\\
 \hline SISO & single-input single-output\\
 \hline SS-OFDMA & \begin{tabular}{c} spatial-spread orthogonal\\ frequency-division multiple-access \end{tabular}\\
 \hline SFFT & symplectic finite Fourier transform\\
 \hline STO & symbol-timing offset\\
 \hline Tbps & terabits-per-second\\
 \hline THz & terahertz\\
 \hline TIV & time-invariant\\
 \hline TTI & transmit time interval\\
 \hline TV & time-variant\\
 \hline UFMC & universal filtered multi-carrier\\
 \hline UM-MIMO & ultra-massive multiple-input multiple-output\\
 \hline UMMTC & ultra-massive machine-type communications\\
 \hline WOLA-OFDM & windowed overlap-and-add OFDM\\
 \hline ZF & zero-forcing\\
 \hline
\end{tabular}
\label{table:AbbreviationsTable2}
\vspace{-6mm}
\end{table}

\bibliographystyle{IEEEtran}
\bibliography{IEEEabrv,bibliography}

\end{document}